%% file: Maxwell.tex
\documentclass[11pt]{elsarticle}

\usepackage{amssymb}
\usepackage{amsmath}
\usepackage{amsfonts}
\usepackage{graphicx}
\usepackage{color}
\usepackage{amsmath,amsthm,amsfonts,epsfig,setspace}
\numberwithin{equation}{section}

\usepackage{multicol}
\usepackage{stmaryrd}
\usepackage{pgf}
\usepackage{pgffor}
\usepgflibrary{plothandlers}
\usepackage{pgfplots}
\pgfplotsset{compat=newest}
 \pgfplotsset{width=15cm}
\pgfplotsset{plot coordinates/math parser=false}
\newlength\figureheight
\newlength\figurewidth

\newcommand{\ds}{\displaystyle}
\def\nm{\noalign{\medskip}}
\newtheorem{theorem}{Theorem}[section]
\newtheorem{remark}{Remark}[section]
\newtheorem{corollary}{Corollary}[section]
\newtheorem{definition}{Definition}[section]
\newtheorem{lemma}{Lemma}[section]

\newtheorem{proposition}{Proposition}[section]

\newtheorem{example}{Example}

 \def\dr{\partial}
\def \Vh0{\stackrel{\circ}{V}_h}

   \def\eps{\varepsilon}

\def\l|{\left|}
\def\r|{\right|}

\newcommand{\cqfd}{\hfill $\square$\\ \medskip}
\newcommand{\R}{\mathbb{R}}
\newcommand{\N}{\mathbb{N}}

\newcommand{\C}{\mathbb{C}}

\newcommand{\nubf}{\boldsymbol{\nu}}
\newcommand{\phibf}{\boldsymbol{\varphi}}
\newcommand{\psibf}{\boldsymbol{\psi}}

\newcommand{\fbf}{\mathbf{f}}

\newcommand{\ebf}{\mathbf{e}}
\newcommand{\gbf}{\mathbf{g}}

\newcommand{\dbf}{\mathbf{d}}

\newcommand{\Ebf}{\mathbf{E}}
\newcommand{\Hbf}{\mathbf{H}}

\newcommand{\dd}{\mathrm{d}}
\newcommand{\tin}{\text{ in }}
\newcommand{\ton}{\text{ on }}

\definecolor{tagada}{RGB}{235,23,55} 
\definecolor{vertperso}{RGB}{0,125,0} 

\title{Shape and size dependence of dipolar plasmonic resonance of nanoparticles}
\author[rvt]{Habib Ammari }
\ead{habib.ammari@math.ethz.ch}
\author[langevin]{Pierre Millien \corref{cor1}}
\ead{pierre.millien@espci.fr}

\cortext[cor1]{Corresponding author}

\address[rvt]{Department of Mathematics, ETH Z\"urich, 
R\"amistrasse 101, CH-8092 Z\"urich, Switzerland}
\address[langevin]{Institut Langevin,  1 Rue Jussieu, 75005 Paris, France}
\begin{document}

\maketitle

\begin{abstract}
The aim of this paper is to present a new approach, based on singular volume integral equations, in order to compute the size dependency 
of plasmonic resonances. The paper also provides rigorous derivations of the extinction and absorption cross sections for elliptical particles.
 
\end{abstract}

\def\keywords2{\vspace{.5em}{\textbf{  Mathematics Subject Classification
(MSC2000).}~\,\relax}}
\def\endkeywords2{\par}
\keywords2{35R30, 35C20.}

\def\keywords{\vspace{.5em}{\textbf{ Keywords.}~\,\relax}}
\def\endkeywords{\par}
\keywords{plasmonic resonance, volume integral equation, singular integrals.}

\section{Introduction}
\subsection{Position of the problem}
The optical properties of metallic nanoparticle have been a subject of great interest in the past decades. They have the ability to exhibit \emph{plasmonic resonances}, which are strong enhancement of the scattering and absorption cross sections at certain frequencies. This capacity to interact strongly with light is a key to many major innovations in nanophotonics \cite{lukas,solar, sarid}, in biomedical imaging \cite{jain, link}, cancer treatment therapy \cite{baffou}. For a nice review of some of these applications we refer the reader to \cite{dreaden2012golden}.

These resonances have been theoretically and experimentally studied by the physics community.
It has been experimentally shown \cite{link1999size} (via measurements of the extinction and absorption cross sections) and numerically (simulations of the Maxwell equations, often via a \emph{coupled dipoles} method, see \cite{kelly2003optical,hao2004synthesis}) that the frequency at which a metallic nanoparticle resonates depends on\begin{enumerate}
\item[(i)] the shape of the particle;
\item[(ii)] the type of metal;
\item[(iii)] the surrounding medium;
\item[(iv)] the size of the particle.
\end{enumerate}
Plasmonic resonances have been the subject of some theoretical work as well in the physics community.
In the case of a spherical particle, the classical Mie theory explains points $(ii)$ to $(iv)$.  In the case where the particle is not spherical, using the \emph{quasi-static approximation} and solving Laplace equation, computations of the polarizability for some simple shapes have given a lot of insights on points $(i)$, $(ii)$, and $(iii)$; see, for instance, \cite{ellipse}. Moreover, the conservation of energy fails in the quasi-static theory, due to the absence of radiative loss. This issue has been dealt by adding a radiative correction \cite{albaladejo2010radiative}.

The size dependence has been more problematic. Some corrections of the quasi-static approximation, sometimes called the \emph{modified 
long-wavelength approximation}, or computations of a \emph{dynamic polarizability} have tackled this issue \cite{meier1983enhanced,schatz1984theoretical,kelly2003optical,moroz2009depolarization}. Nevertheless, they heavily rely on strong assumptions and are valid only for spheroidal shapes.

In the mathematical community, plasmonic resonances are a more recent subject of interest.  In the quasi-static approximation, plasmonic resonances were shown to be an eigenvalue problem linked to the Neumann Poincar\'e operator \cite{grieser2014plasmonic, HK1,HK4}. It was then showed that Maxwell's equation yields a similar type of eigenvalue problems, and a computation of the polarizability for small plasmonic particle was given, solving items $(i)$ to $(iii)$ for a general regular shape \cite{ammari2016surface}. Note that these studies were all done in the case where the shape of the particle is assumed to have some regularity, and the theory breaks down when the particle has corners. Some recent progress has been made on this topic \cite{bonnetier2017characterization, HK3,perfekt2017essential}.

The size dependance has been justified in \cite{ammari2015mathematical,HK2} in the scalar case (transverse electric or transverse magnetic) and in \cite{ammari2016mathematicalruiz} in the Maxwell setting.  However, practical computations of this size dependency remains complicated. We aim here at presenting a new approach, based on a singular volume integral equation, to compute this size dependency. Our integral volume approach can be extended to the case where the shape of the particle has corners.

\subsection{Main contribution}\label{sec:maincontrib}
In this work, using a volume integral equation,  we show that the resonant frequencies at which a nanoparticle  of characteristic size $\delta$ exhibits plasmonic resonances occurs can be written as a nonlinear eigenvalue problem:
\begin{align}\label{eq:general}
\mathrm{Find}\ \omega\  \mathrm{such \ that \ } f(\omega)\in \sigma\left(\mathcal{T}^{(\omega \delta)}\right)
\end{align}
for some nonlinear function $f$ and some operator $\mathcal{T}^{(\omega \delta)}$ (see Definition \ref{de:plasmonic}).

These types of problems are extremely difficult to handle in their generality. In this work, we add some assumptions arising from experimental observations and classical electromagnetic theory to compute solutions of  \eqref{eq:general} in a regime that corresponds to practical situations.

The perturbative analysis presented in this work is based on the following assumptions:
\begin{enumerate}
\item[(i)] The size $\delta$ of the particle is small compared to the wavelenght of the lights at plasmonic frequencies: \begin{align*}
\delta \frac{\omega}{c} \ll 1;
\end{align*}
\item[(ii)] The particle is constituted of metal, whose permittivity can be described by a Drude-Lorentz type model \cite{ordal1983optical}.
\end{enumerate}

In this regime, we show that 
(Theorem \ref{theo:spectrumperturbation}):
\begin{align*}
\frac{\partial}{\partial \omega} \sigma(\mathcal{T}^{(\omega\delta)}) \sim \frac{\delta}{c} \ll 1.
\end{align*} 

And using that we give the following procedure for solving \eqref{eq:general}:
\begin{itemize}
\item Find $\omega_0$ such that $f(\omega_0) \in \sigma(\mathcal{T}^{(0)})$;
\item Compute $\sigma(\mathcal{T}^{(\delta\omega_0)})$ by a perturbative method;
\item Find $\omega_1$ such that $f(\omega_1)\in \sigma (\mathcal{T}^{(\delta \omega_0)})$.
\end{itemize}
Since, in practical situations $\frac{\partial}{\partial \omega} f(\omega) \gg \frac{\delta}{c}$ (this comes from the fact that the particle is metallic and can be checked numerically, see Appendix \ref{sec:justif} for more details), one can see that $\omega_1$ is a good approximated solution of problem \eqref{eq:general}.

\subsection{Additional contributions}
 In this paper, we also show that in the case where the particle has an elliptic shape, the dipole resonance of the nanoparticle (and its dependence on the size of the particle) can be very easily computed using the $\mathbf{L}$ dyadic that can be found in the physics literature \cite{van1991singular,yaghjian1980electric}. 
This dyadic $\mathbf{L}$ is often incorrectly derived in the literature. In Appendix \ref{appendix:singular} we give a correct derivation of $\mathbf{L}$, as well as some precisions on some common misconceptions about singular integrals found in the classical literature on electromagnetic fields.
We also give formulas for the computations of some observable quantities such as the extinction and absorption cross sections for elliptical particles (see Section \ref{sec:observableelliptic}). To the best of our knowledge,  this is the first time that a formal proof is given for these type of computations.

\section{Model and definition}

\subsection{Maxwell's equations}
\begin{figure}[!h]
\def\svgwidth{0.7\linewidth}
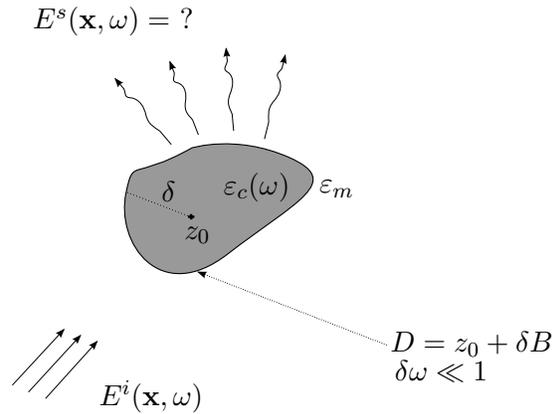
\caption{Schematic representation of the scattering problem.}
\end{figure}
We consider the scattering problem of a time-harmonic wave incident on a plasmonic nanoparticle. Denote by $\varepsilon_0$ and $\mu_0$ the electric permittivity and the magnetic permeability of the vacuum and by $c_0=(\varepsilon_0 \mu_0)^{-1/2}$ the speed of light in the vacuum.  The homogeneous medium is characterized by its relative electric permittivity $\varepsilon_m$ and relative magnetic permeability $\mu_m$, while the particle occupying a bounded and simply connected domain of center of mass $z_0$: $$D=z_0+\delta B \Subset\mathbb{R}^3$$ with $\mathcal{C}^{1,\alpha}$ boundary  is characterized by its electric permittivity $\varepsilon_c$ and its magnetic permeability $\mu_c$, both of which may depend on the frequency. We  assume that $\Re \eps_c <0, \Im \eps_c >0$ and define
\begin{align}\label{eq:defkm}
k_m = \frac{\omega}{c_0} \sqrt{\eps_m \mu_m}, \quad k_c =\frac{\omega}{c_0} \sqrt{\eps_c \mu_c},
\end{align}
and
\begin{align}
\eps_D(\omega)= \eps_m \chi(\R^3 \backslash \bar{D}) + \eps_c(\omega) \chi(\bar{D}), \quad
\mu_D= \mu_m \chi(\R^3 \backslash \bar{D})+ \mu_c \chi({D}),
\end{align}
where $\chi$ denotes the characteristic function.  We assume that the particle is nonmagnetic, i.e.,  $\mu_m=\mu_c$. Throughout this paper, we assume that $\eps_m$ is real and strictly positive and that $\Re k_c <0$ and $\Im k_c >0$.

For a given plane wave solution $(\Ebf^i, \Hbf^i)$ to the Maxwell
equations
\begin{equation*}
 \left \{
 \begin{array}{ll}
\nabla\times{\Ebf^i} = i\omega\mu_m{\Hbf^i} \quad &\mbox{in } \mathbb{R}^3,\\
\nabla\times{\Hbf^i} = -i\omega\varepsilon_m {\Ebf^i}\quad &\mbox{in
} \mathbb{R}^3,
 \end{array}
 \right .
 \end{equation*}
let $(\Ebf,\Hbf)$ be the solution to the following Maxwell equations: 
\begin{equation}
\label{eq:maxwell} \left\{
\begin{array}{ll}
 \nabla \times \Ebf = i \omega \mu_D \Hbf &  \mbox{in} \quad \mathbb{R}^3\setminus \dr D, \\
 \nabla  \times \Hbf= - i \omega \varepsilon_D \Ebf & \mbox{in} \quad \R^3\setminus \dr D, \\
 {[}{\nubf} \times \Ebf]= [{ \nubf} \times \Hbf] = 0 & \mbox{on} \quad \dr D,
\end{array}
\right.
\end{equation}  subject to the Silver-M\"{u}ller radiation condition:
 $$\lim_{|x|\rightarrow\infty} |x| (\sqrt{\mu_m} (\Hbf- \Hbf^i) \times\hat{x}-\sqrt{\varepsilon_m} (\Ebf-\Ebf^i))=0,$$
 where $\hat{x} = x/|x|$. Here, $[{\nubf} \times{\Ebf}]$ and $[{\nubf} \times{\Hbf}]$ denote the jump of ${\nubf} \times{\Ebf}$
and ${\nubf} \times{\Hbf}$ along $\dr D$, namely,
 \begin{equation*}
 [\nubf\times{\Ebf}]=(\nubf\times \Ebf)\bigr|_+ -(\nubf \times \Ebf)\bigr|_-,\quad [\nubf\times{\Hbf}]=(\nubf\times \Hbf)\bigr|_+ -(\nubf\times \Hbf)\bigr|_-.
  \end{equation*} 

\begin{proposition} If $\mathcal{I}\left[\frac{\varepsilon_c}{\varepsilon_m}\right]\neq 0$ , then problem (\ref{eq:maxwell}) is well posed.
Moreover, if we denote by $(\Ebf,\Hbf)$ its unique solution, then $(\Ebf,\Hbf)\big\vert_D \in H(\mathrm{curl}, D)$ and $(\Ebf,\Hbf)\big\vert_{\R^3\setminus D} \in H_{\mathrm{loc}}(\mathrm{curl},\R^3\setminus \overline{D})$.

\end{proposition}
\proof The well-posedness is addressed in \cite{torres1998maxwell,costabel2012essential,ammari2015mathematical}.
\cqfd


We also denote by $G^{k_m}$ the scalar outgoing Green function for the homogeneous medium, i.e., the unique solution in the sense of distributions of
\begin{equation}\label{eq:defGkm}
\left(\Delta +\frac{\omega^2}{c_0^2}\varepsilon_m\mu_m\right)G^{k_m}(\cdot,z)=\delta_z \quad \tin \R^3,
\end{equation} subject to the Sommerfeld radiation condition. $G^{k_m}$ is given by (see \cite{nedelec2001acoustic}): 
\begin{align}\label{eq:valueGkm}
G^{k_m}(x,z)=&\frac{e^{ik_m\vert x-z\vert}}{4\pi\vert x-z\vert}. 
\end{align}

\subsection{Volume integral equation for the electric field}
We start by defining a singular integral operator, sometimes known as the \emph{magnetization integral operator} \cite{friedman1984spectral}. 
\begin{definition} Introduce

\begin{align*} 
\mathcal{T}_D^k: \begin{aligned} L^2(D,\R^2)& \longrightarrow L^2(D,\R^2) \\
\mathbf{f} &\longmapsto k^2 \int_{D} G^k(x,y) \fbf(y)\dd y -\nabla \int_D \nabla G^k (\cdot,y) \cdot \mathbf{f}(y) \dd y .
\end{aligned}
\end{align*}
\end{definition}

We then give the equation satisfied by the electric field:

\begin{proposition} \label{prop:LSequation} 
The electric field inside the particle satisfies the volume integral equation 
(or Lippmann-Schwinger equation):
\begin{align}\label{eq:fieldinside}
\left(\frac{\varepsilon_m}{\varepsilon_m - \varepsilon_c} I -  \mathcal{T}^k_D\right) \Ebf = \frac{\varepsilon_m}{\varepsilon_m - \varepsilon_c} \Ebf^i. 
\end{align}

\end{proposition}
\proof See \cite[Chapter 9]{colton2012inverse} or \cite{costabel2010volume}.

\subsection{Plasmonic resonances as an eigenvalue problem}

\begin{definition}\label{de:plasmonic}
We say there is a plasmonic resonance if $$\frac{\varepsilon_c}{\varepsilon_m-\varepsilon_c} \in \sigma \left(\mathcal{T}_D^k\right).$$
\end{definition}

\subsection{Dipole resonance}\label{sec:dipole}\begin{definition}\label{de:dipole}
The dipole moment of a particle is given by
\begin{align*}
\mathbf{P}=\int_D\mathbf{p}(x) \dd x = \int_D \varepsilon_m \chi(x) \Ebf(x)\dd x = \int_D (\varepsilon_c-\varepsilon_m) \Ebf(x) \dd x. 
\end{align*}
\end{definition}

We say that there is a dipolar plasmonic resonance if the dipole moment $\mathbf{P}$ satisfies $$ |\mathbf{P} | \gg  \big| (\varepsilon_c-\varepsilon_m) \int_D \Ebf^i(x) \dd x \big| .$$ Therefore, we want to compute the values of  $\varepsilon_c$ and $\varepsilon_m$ such that

\begin{enumerate}
\item[(i)] \begin{align*}
\lambda := \frac{\varepsilon_c}{\varepsilon_m-\varepsilon_c} \in \sigma \left(\mathcal{T}_D^k\right);
\end{align*}
\item[(ii)] One of the eigenvectors $\phibf_\lambda$ associated with $\lambda$ has non zero average:
\begin{align*}
\frac{1}{\vert D\vert}\int_D \phibf_\lambda \not= 0.
\end{align*}
\end{enumerate}

\section{The quasi-static approximation}
In this section, we study the case when the particle has finite size $\delta\neq 0$ and $\delta k_m \ll 1$. This corresponds to the usual \emph{quasi-static approximation}. It has already been shown in \cite{ammari2016surface,ammari2015mathematical} that the solution of Maxwell's or Helmholtz equation converge uniformly when $\delta k_m \rightarrow 0$ to the solution of the quasi-static problem in the case of negative index materials.

\begin{proposition}
In the quasi-static approximation, the excitation field $\Ebf^i$ becomes constant, the electric field can be written as the gradient of a potential $\Ebf = \nabla u$ and the scattering problem described by  (\ref{eq:fieldinside}) becomes:
\begin{align*}
\text{Find } u \ \text{such that \ } \left\{\begin{aligned} \nabla \cdot \left(  \varepsilon_D(x) \nabla u \right)= 0,  \\
u(x) -\Ebf^i \cdot x = O\left( \vert x \vert^{-1}\right) . \end{aligned}\right. 
\end{align*}
Equivalently,  $\Ebf = \nabla u$ is a solution of the following integral equation:
\begin{align}\label{eq:static}
\left(\frac{\varepsilon_m}{\varepsilon_m - \varepsilon_c} I -  \mathcal{T}^0_D\right) \Ebf = \frac{\varepsilon_m}{\varepsilon_m - \varepsilon_c} \Ebf^i.
\end{align}
\end{proposition}
\begin{remark}These types of transmission / exterior problems have been extensively treated in the literature. For more details on the well posedness, the appropriate functional spaces, and the study of small conductivity inhomogeneities we refer to \cite{nedelec2001acoustic,ammari2004reconstruction}.
\end{remark}
\begin{proposition}Let $y=z_0 +\delta\tilde{y}$ and write $\tilde{u}(\tilde{y})=u(y)$, and $\tilde{u}^i(\tilde{y}) = u^i(y)$. Then $\tilde{u}$ solves:
\begin{align*}
\left(\frac{\varepsilon_c}{\varepsilon_m - \varepsilon_c} I -  \mathcal{T}^0_B\right) \nabla \tilde{u} = \frac{\varepsilon_c}{\varepsilon_m - \varepsilon_c} \nabla \tilde{u}^i. 
\end{align*}
\end{proposition}
\proof This is a direct consequence of Theorem \ref{theo:changeofvariable}.

\subsection{Spectral analysis of the static operator, link with Neumann-Poincar\'e operator} It has been shown in \cite{ammari2016surface,ammari2015mathematical} that the plasmonic resonances are linked to the eigenvalues of the Neumann-Poincar\'e operator. In this subsection, we show that the surface integral approach and the volume integral approach are coherent. The link between the volume integral operator and the Neumann Poincar\'e operator is summed up in Corollary \ref{cor:linkk*}.
We first recall the definition of the Neumann-Poincar\'e operator
\begin{definition}\label{de:NPoperator}
The operator $\mathcal{K}^*_{D}: L^2(\partial D) \rightarrow L^2(\partial D)$ is defined by
\begin{equation}
    \mathcal{K}^*_{ D} [\varphi] (x) := \frac{1}{\omega_d} \int_{\partial D} \frac{( x-y)\cdot \nubf(x)  }{|x-y|^d} \varphi(y) d \sigma(y) \,, 
    \label{operatorK}
\end{equation}
with $\nubf(x)$ being 
the outward normal at $x \in \partial D$, $\omega_d$ the measure of the unit sphere in dimension $d$, and  $\sigma$ the Lebesgue measure on $\partial D$. We note that $\mathcal{K}^*_{D}$ maps $L^2_0(\partial D)$ onto itself (see, for instance, \cite{ammari2004reconstruction}).
\end{definition}

Recall the orthogonal decomposition
\begin{align*}
L^2(D,\R^3) = \nabla H^1_0(D) \oplus \mathbf{H}(\text{div\ } 0,D) \oplus \mathbf{W},
\end{align*}
where $\mathbf{H}(\text{div\ } 0,D) $ is the space of divergence free $L^2$ vector fields and $\mathbf{W}$ is the space of gradients of harmonic $H^1$ functions.
We start with the following result from \cite{costabel2012essential}:

\begin{proposition}\label{prop:decompT0}

The operator $\mathcal{T}_D^0$ is a bounded self-adjoint map on $L^2(D,\R^2)$ with $  \nabla H^1_0(D)$, $\mathbf{H}(\text{div\ } 0,D)$ and $\mathbf{W}$ as invariant subspaces. On $\nabla H^1_0(\Omega)$, $\mathcal{T}_D^0[\phibf]=\phibf$, on $\mathbf{H}(\text{div\ } 0,D)$, $\mathcal{T}_D^0[\phibf]=0$ and on $\mathbf{W}$: $$\nubf\cdot \mathcal{T}_D^0[\phibf]= \left(\frac{1}{2} + \mathcal{K}_D^*\right)[\phibf\cdot \nubf] \ton \partial D . $$
\end{proposition}
\proof
The proof can be found in \cite{friedman1984spectral, costabel2012essential}. \cqfd

From this, it immediately follows that the following corollary holds. 
\begin{corollary}\label{cor:linkk*}
\noindent Let $\lambda \not=1$. Let $\phibf_D\not\equiv 0$ be such that
\begin{align*}
\lambda \phibf_D - \mathcal{T}_D^0[\phibf_D]= 0 \quad \tin \ D.
\end{align*}
Then,
\begin{align*}
&\phibf_D\in \mathbf{W},\\
&\nabla \cdot \phibf _D = 0 &\tin D, \\
&\lambda \phibf_D = \nabla \mathcal{S}_D[\phibf_D \cdot \nubf]&\tin D , \\
&\lambda \phibf_D \cdot \nubf = \left(\frac{1}{2} + \mathcal{K}_D^*\right)[\phibf_D\cdot \nubf] &\ton \partial D . 
\end{align*}
Letting $u_D=\mathcal{S}_D[\phibf_D\cdot\nubf]$, we have:
\begin{align}\label{eq:laplacien}
\left\{ \begin{aligned}
&\Delta u_D= 0 &\tin \R^3\setminus \partial D ,\\
&[\partial_n u_D ]=\phibf_D\cdot\nubf &\ton \partial D .
\end{aligned}\right.
\end{align}
\end{corollary}

\begin{proposition}
If the boundary of $D$ is $\mathcal{C}^{1,\alpha}$, then  $\mathcal{T}_D^0\big\vert_\mathbf{W}$ : \mbox{$\mathbf{W} \longrightarrow \mathbf{W} $} is a compact operator.
\end{proposition}
\proof The operator $\mathcal{T}_D^0$ is a bounded map from $\mathbf{W}$ to $H^1(D,\R^3)$ \cite{friedman1984spectral, mikhlin2014multidimensional}. The $\mathcal{C}^{1,\alpha}$ regularity of $\partial D$ and the usual  Sobolev embedding theorems ensure its compactness (see \cite[Chapter $9$]{brezis2010functional}).

\begin{proposition}\label{prop:eigenbasis} The set of eigenvalues $(\lambda_n)_{n\in \N}$ of $\mathcal{T}^0_D\big\vert_\mathbf{W}$ is discrete, and the associated eigenfunctions $(\phibf_n)$  form a basis of $\mathbf{W}$. We have:
\begin{align*}
\mathcal{T}_D^0\big\vert_\mathbf{W} = \sum_n \lambda_n \langle \phibf_n, \cdot \rangle \phibf_n . 
\end{align*}
\end{proposition}

\subsection{Link between $\sigma(\mathbf{L}_B)$ and $\sigma(\mathcal{T}_B^0)$, and the dipole resonances for an ellipse}\label{sec:staticellipse}
As explained in Section \ref{sec:dipole}, to understand the dipole resonances of the particle, we need to compute the eigenvectors of $\mathcal{T}_B^0$ that have a non zero average, i.e.,  that are not orthogonal in the $L^2(B)$ sense to every $\ebf_i$, $i \in \{1,2,3\}$, where $(\ebf_1,\ebf_2,\ebf_3)$ is an orthonormal basis of $\mathbb{R}^3$. 

In general, constant vector fields over $B$ are not eigenvectors of $\mathcal{T}_B$.
Nevertheless in the case where $B$ is an ellipse, then constant vector fields can be eigenvectors for $\mathcal{T}_B$.
This is essentially a corollary of \emph{Newton's shell theorem}.

\begin{theorem}\label{theo:ellipseconstant}
If $B$ is an ellipse centered at the origin, then the following holds:
Let $\phibf \in L^2(B,\R^3)$ and let $\lambda \in \R\setminus \{0,1\}$ be such that
\begin{align*}
\left\{\begin{aligned} \lambda \phibf- \mathcal{T}_B^0 [\phibf]= 0 , \\
\int_B \phibf \not = 0.
\end{aligned}\right.
\end{align*}
Then, 
\begin{align*}
 \left(\lambda I - \mathbf{L}_B\right)\int_B \phibf_0= 0.
\end{align*}
\end{theorem}
\begin{remark}The operator we are considering is essentially the double derivative of a classical Newtonian potential. When the domain is an ellipse, the Newtonian potential of a constant is a second order polynomial. Therefore, its second derivative is a constant. Hence, the possibility to have constant eigenvectors for $\mathcal{T}_B$ occurs. This property characterizes ellipses. In fact, it is the weak Eshelby conjecture;  see \cite{kang2008solutions} for more details.
\end{remark}

\proof 
For the proof we need the following lemma from \cite{di2016newtonian}:
\begin{lemma}\label{lem:ellipseconstant}
If $B$ is an ellipse, then, for any $\phibf_0 \in \R^3$, 
\begin{align*}
\mathcal{T}_B[\phibf_0]=\mathbf{L}_B \phibf_0.
\end{align*}
\end{lemma}
Combining this with Proposition \ref{prop:eigenbasis}, and using the orthogonality between the eigenvectors of $\mathcal{T}_B^0$, one gets the result. \cqfd
\begin{corollary}\label{cor:ellipsconstant}
Let $\Ebf\in \R^3\setminus\left\{ 0 \right\}$ be such that $ \lambda \Ebf= \mathbf{L}_B \Ebf$. 
Then,
\begin{align*}
\mathcal{T}^0 [\Ebf]= \lambda \Ebf. 
\end{align*}
\end{corollary}
\proof This is a direct consequence of Lemma \ref{lem:ellipseconstant}.

\subsection{Static polarizability of an ellipse}
In this subsection,  we assume that $B$ is an ellipse.
\begin{definition}
The polarizability $\mathcal{M}$ is the matrix linking the average incident electrical field to the induced dipolar moment. It is defined by
\begin{align*}
\mathbf{p}=\mathcal{M} \left( \frac{1}{\vert D \vert }\int_D \Ebf^i\right).
\end{align*}

\end{definition}

\begin{theorem}\label{theo:staticpolar} The static polarizability $\mathcal{M}$ of the particle $B$ is given by 

\begin{align}\label{eq:alphastatic}
 \mathcal{M} =  \delta^3 \varepsilon_0\left(\varepsilon_c -1 \right) \left(\frac{\varepsilon_m}{\varepsilon_m - \varepsilon_c} + \mathbf{L}_B\right)^{-1}. 
\end{align}
\end{theorem}
\begin{remark}
The polarizability is used to compute different observables such as the scattering and extinction cross sections of the particle (see Section \ref{sec:observableelliptic}).
\end{remark}

\proof
We recall equation \eqref{eq:static} for the electric field inside the particle
\begin{align*}
\left(\frac{\varepsilon_m}{\varepsilon_m - \varepsilon_c} I -  \mathcal{T}^0_D\right) \nabla u = \frac{\varepsilon_m}{\varepsilon_m - \varepsilon_c} \nabla u^i.
\end{align*}
We now remark that the operator
\begin{align*}
\mathcal{P}: L^2(D,\R^3) \longrightarrow & L^2(D,\R^3) \\
\fbf \longmapsto& \int_D \fbf
\end{align*}
is the projector onto the subspace of $L^2$-spanned by constant functions. 
Compose the previous equation with $\mathcal{P}$ we get:
\begin{align*}
\mathcal{P} \circ \left(\frac{\varepsilon_c}{\varepsilon_m - \varepsilon_c} I -  \mathcal{T}^0_D\right) [ \nabla u ] = \mathcal{P}[\nabla u^i].
\end{align*}
We now use Proposition \ref{prop:eigenbasis} to diagonalise $\mathcal{T}^0$:
\begin{align*}
\mathcal{P} \left(\sum_i \left(\frac{\varepsilon_c}{\varepsilon_m - \varepsilon_c} + \lambda_i\right) \left\langle \phibf_n, \nabla u  \right\rangle \phibf_n \right)  = \mathcal{P}[\nabla u^i].
\end{align*}
Since the particle is an ellipse, we know by Corollary \ref{cor:ellipsconstant} that $\Ebf_i \in \R^3$, the eigenvectors of $\mathbf{L}_B$ associated with the eigenvalues $\lambda_{\mathbf{L},i}$,  are also eigenvectors for $\mathcal{T}^0$ for some eigenvalues $\lambda_{i_1}$, $\lambda_{i_2}$ and $\lambda_{i_3}$ . Moreover, $\lambda_{i_1}= \lambda_{\mathbf{L},1}$, $ \lambda_{i_2}= \lambda_{\mathbf{L},2}$ and $ \lambda_{i_3}= \lambda_{\mathbf{L},3}$
We can also note that $(\Ebf_1, \Ebf_2, \Ebf_3)$ span the image of the projector $\mathcal{P}$. By orthogonality of the eigenvectors of $\mathcal{T}^0$, we obtain that
\begin{align*}
\mathcal{P} \left(\sum_i \left(\frac{\varepsilon_c}{\varepsilon_m - \varepsilon_c} + \lambda_i\right) \left\langle \phibf_n, \nabla u  \right\rangle \phibf_n \right) =  \sum_i \left(\frac{\varepsilon_m}{\varepsilon_m - \varepsilon_c} + \lambda_{\mathbf{L},i}\right) \left\langle \Ebf_i, \nabla u  \right\rangle \Ebf_i 
\end{align*}
Noticing that $\langle \Ebf_i , \nabla u \rangle = \langle \Ebf_i ,\mathcal{P}(\nabla u) \rangle = \Ebf_i \cdot \mathcal{P}(\nabla u)$, 
we obtain that the right-hand side of the previous equation is exactly the expression of $$\left(\frac{\varepsilon_c}{\varepsilon_m - \varepsilon_c} + \mathbf{L}_B\right)\mathcal{P}(\nabla u) $$ in the basis $(\Ebf_1, \Ebf_2)$. Therefore, we have shown that 
\begin{align*}
\mathcal{P} \circ \left(\frac{\varepsilon_c}{\varepsilon_m - \varepsilon_c} I -  \mathcal{T}^0_D\right) [ \nabla u ] = \left(\frac{\varepsilon_c}{\varepsilon_m - \varepsilon_c} + \mathbf{L}_B\right)\mathcal{P}(\nabla u).
\end{align*}
Thus
\begin{align*}
\int_D \nabla u = \left(\frac{\varepsilon_c}{\varepsilon_m - \varepsilon_c} + \mathbf{L}_B\right)^{-1} \int_D \nabla u^i.
\end{align*}
Using Definition \ref{de:dipole} of the induced dipole moment,  we get the result.
\cqfd

\subsection{Static polarizability of an arbitrary particle}

In the case where the particle occupies an arbitrary $\mathcal{C}^{1,\alpha}$ domain, the volume integral approach does not yield a simple expression for the polarizability. Nevertheless, the layer potential approach gives the well known polarization tensor.
The validity of the polarization tensor formula for negative index material has been shown in \cite{ammari2015mathematical,ammari2016surface}.

We recall here the formula for completeness:

\begin{theorem}{\cite{ammari2015mathematical,ammari2016surface}} ~
The static polarizability is given by
\begin{align*}
\alpha = \delta^3\varepsilon_0(\varepsilon -1) \int_{\partial B}y\left(\frac{ \varepsilon+1}{2( \varepsilon-1)} I - \mathcal{K}_B^*\right)^{-1}[\nubf](y) \dd \sigma(y), 
\end{align*}where $\varepsilon= \frac{\varepsilon_c}{\varepsilon_m}$.
\end{theorem}

\section{Perturbative approach: spectral analysis of the dynamic operator}
\label{sec:perturbative}

In this section,  we aim at finding $\tilde{\lambda}$ such that there exists some $\mathbf{f}\not\equiv 0\in L^2(B,\R^3)$ such that
\begin{align*}
\left(\tilde{\lambda} I - \mathcal{T}^{\delta k}_B\right) [\mathbf{f}] =0.
\end{align*}

 Let $\lambda_{n_0}$ be an eigenvalue of $\mathcal{T}^0$.  Let $V\subset \C$ be a neighborhood of $\lambda_{n_0}$ such that $\lambda I - \mathcal{T}^0$ is invertible for every $\lambda \in V$. Let $\phibf_{n_0} \in L^2(B,\R^3)$ be a unitary eigenvector associated with $\lambda_{n_0}$.
\begin{lemma}\label{lem:decomposition}For any $\lambda \in V$, the following decomposition holds: \begin{align*}
\left(\lambda I - \mathcal{T}^0\right)^{-1} = \frac{\langle \phibf_{n_0}, \cdot \rangle }{\lambda - \lambda_{n_0}} \phibf_{n_0} + \mathcal{R}(\lambda),
\end{align*}
where 
\begin{align*}
\begin{aligned}\C \longrightarrow &\left( L^2(B,\R^3) \rightarrow L^2(B,\R^3)\right) \\
\lambda  \longmapsto & \mathcal{R}(\lambda)
\end{aligned}
\end{align*} is holomorphic in $\lambda$.
\end{lemma}
\proof Denote by $\mathcal{P}_1 : L^2(B,\R^3)\rightarrow L^2(B,\R^3)$ and $\mathcal{P}_2: L^2(B,\R^2)\rightarrow L^2(B,\R^3)$  the orthogonal projections on $\nabla H^1_0(B)$ and $\mathbf{H}(\text{div\ }0, B)$, respectively.
Using Propositions \ref{prop:decompT0} and  \ref{prop:eigenbasis}, we can write:
\begin{align*}
\lambda I - \mathcal{T}^0 = \sum (\lambda-\lambda_n) \langle \phibf_n, \cdot \rangle \phibf_n + (\lambda -1) \mathcal{P}_1 + \lambda \mathcal{P}_2.
\end{align*}
The result immediately follows.\cqfd

\begin{lemma}\label{lem:oneeigenvalue}
Let $\lambda_{n_0}$ be an eigenvalue for $\mathcal{T}^0$. Then, if $\vert k\vert $ is small enough, there exists a neighborhood $V\subset \C$ of $\lambda_{n_0}$  such that $\mathcal{T}^{\delta k}_B$ has exactly one eigenvalue in $V$.
\end{lemma}
\proof We start by writing:
\begin{align*}
\lambda I - \mathcal{T}^{\delta k} = \lambda I - \mathcal{T}^0 + \left(\mathcal{T}^0 - \mathcal{T}^{\delta k}\right).
\end{align*}
Recall that there exists $V\subset \C$ such that $\lambda I -\mathcal{T}^0$ is invertible for every $\lambda \in V\setminus\{ \lambda_{n_0}\}$.
Therefore, for $\lambda \in V\setminus\{ \lambda_{n_0}\}$, 
\begin{align*}
\lambda I - \mathcal{T}^{\delta k} = \left(\lambda I - \mathcal{T}^0 \right)\left( I + \left(\lambda I - \mathcal{T}^0 \right)^{-1}\left(\mathcal{T}^0 - \mathcal{T}^{\delta k}\right)\right).
\end{align*}
Using Lemma \ref{lem:decomposition}, we get
\begin{align*}
\left(\lambda I - \mathcal{T}^{\delta k}\right)[\fbf]= \fbf +  \frac{\langle \phibf_{n_0} ,\left(\mathcal{T}^0 - \mathcal{T}^{\delta k}\right)[\fbf]\rangle  }{\lambda - \lambda_{n_0} } \phibf_{n_0} + \mathcal{R}(\lambda) \left(\mathcal{T}^0 - \mathcal{T}^{\delta k}\right)[\fbf].
\end{align*}
We can show that \begin{align*}
\left\Vert \mathcal{T}^0 - \mathcal{T}^{\delta k}\right\Vert \longrightarrow 0 \quad( \delta k \rightarrow 0). 
\end{align*}
Since $\lambda \mapsto \mathcal{R}(\lambda)$ is holomorphic, the compact operator $$\lambda \longmapsto \mathcal{R}(\lambda) \left(\mathcal{T}^0 - \mathcal{T}^{\delta k}\right)$$  converges uniformly to $0$ with respect to $\lambda$ when $k$ goes to $0$.
Since the operator \begin{align*}
\mathcal{K}^k : \begin{aligned} 
 L^2(B,\R^3)& \longrightarrow L^2(B,\R^3) \\
 \fbf &\longmapsto \frac{\langle \phibf_{n_0} ,\left(\mathcal{T}^0 - \mathcal{T}^{\delta k}\right)[\fbf]\rangle  }{\lambda - \lambda_{n_0} } \phibf_{n_0}
\end{aligned}
\end{align*} is a rank one linear operator, the operator 
$I + \mathcal{K}_B^{\delta k}$ is invertible. 

Therefore, there exists $K>0$ such that $\lambda I - \mathcal{T}^k = I + \mathcal{K}^{\delta k} + \mathcal{R}(\lambda) \left(\mathcal{T}^0 - \mathcal{T}^{\delta k}\right)$ is invertible for every $\lambda\in V\setminus \{\lambda_{n_0}\}$ and every $\vert\delta  k \vert  < K$. \cqfd

We can now give an asymptotic formula for the perturbed eigenvalues $\tilde{\lambda}$ of $\mathcal{T}^{\delta k}$:
\begin{proposition}{ The following asymptotic formula for the perturbed eigenvalues holds:}\label{prop:asymptvp}
\begin{align}\label{eq:asymptvp}
\tilde{\lambda}\sim  \lambda_{n_0} - \left\langle\left( \mathcal{T}^0 - \mathcal{T}^{\delta k}\right) \phibf_{n_0}, \phibf_{n_0}\right\rangle_{L^2(B,\R^3)} .
\end{align}
\end{proposition}
\proof
We use the same notations as in the previous lemmas. We have: 
\begin{align*}
\tilde{\lambda} \in \sigma\left(  \mathcal{T}^{\delta k} \right)\cup V\setminus \{\lambda_{n_0}\} \quad  \Leftrightarrow & \quad \exists \fbf \not\equiv 0 \quad \text{such that}\ \left(\tilde{\lambda} I - \mathcal{T}^{\delta k}\right)[\fbf]= 0 \\
\Leftrightarrow & \quad \exists \fbf \not\equiv 0 \quad \text{such that}\ \left( I + \left(\mathcal{T}^0 -\mathcal{T}^{\delta k}\right)\left( \tilde{\lambda}I - \mathcal{T}^0\right)^{-1}\right)[\fbf]= 0 . 
\end{align*}
Using the decomposition stated in Lemma \ref{lem:decomposition} for $\left( \tilde{\lambda}I - \mathcal{T}^0\right)^{-1}$, we get the following equation for $\fbf$ and  $\tilde{\lambda}$:
\begin{align}\label{eq:decompf}
\fbf  + \frac{\langle \phibf_{n_0}, \fbf \rangle}{\tilde{\lambda}-\lambda_{n_0}} \left(\mathcal{T}^0 -\mathcal{T}^{\delta k}\right)[\phibf_{n_0}] +\left(\mathcal{T}^0 -\mathcal{T}^{\delta k}\right)\mathcal{R}(\tilde{\lambda})[\fbf] = 0.
\end{align}
We start by proving that $\langle \phibf_{n_0}, \fbf \rangle\not= 0$. Indeed, if one has $\langle \phibf_{n_0}, \fbf \rangle= 0$, then (\ref{eq:decompf}) becomes
\begin{align*}
\left(I+ \left(\mathcal{T}^0 -\mathcal{T}^{\delta k}\right)\mathcal{R}(\tilde{\lambda})\right)[\fbf]=0.
\end{align*} 
If $k$ is close enough to $0$,  then $\Vert \left(\mathcal{T}^0 - \mathcal{T}^{\delta k}\right)\mathcal{R}(\tilde{\lambda})\Vert <1$ and then  $I+ \left(\mathcal{T}^0 -\mathcal{T}^{\delta k}\right)\mathcal{R}(\tilde{\lambda})$ is invertible and we have $\fbf =0$,  which is a contradiction.

We then note that $\fbf$ and $\frac{\langle \phibf_{n_0}, \fbf \rangle}{\tilde{\lambda}-\lambda_{n_0}} \left(\mathcal{T}^0 -\mathcal{T}^{\delta k}\right)[\phibf_{n_0}]$ are terms of order $O(\vert \fbf\vert)$ whereas the regular part, the term $\left(\mathcal{T}^0 -\mathcal{T}^{\delta k}\right)\mathcal{R}(\tilde{\lambda})[\fbf]$ is of order $O(\delta k \vert \fbf \vert)$. 
We drop the regular part, and take the scalar product against $\phibf_{n_0}$ to obtain that
\begin{align}\label{eq:decompf2}
\langle \phibf_{n_0}, \fbf \rangle + \frac{\langle \phibf_{n_0}, \fbf \rangle}{\tilde{\lambda}-\lambda_{n_0}} \left\langle\left(\mathcal{T}^0 -\mathcal{T}^{\delta k}\right)[\phibf_{n_0}], \phibf_{n_0}\right\rangle = 0.
\end{align}
Finally,  dividing equation (\ref{eq:decompf2}) by $\langle \phibf_{n_0}, \fbf \rangle \neq 0$ yields
\begin{align*}
\tilde{\lambda} = \lambda_{n_0} - \left\langle\left(\mathcal{T}^0 -\mathcal{T}^{\delta k}\right)[\phibf_{n_0}], \phibf_{n_0}\right\rangle.
\end{align*} \cqfd

We also prove the following lemma, giving an approximation of the resolvent of $\mathcal{T}_D^k$ along the direction of the eigenfunction $\phibf_{n_0}$:
\begin{proposition}\label{prop:approxresolv}Let $\gbf\in L^2(B,\R^3)$. If $\fbf\in L^2(B,\R^3)$ is a solution of
\begin{align*}
\left(\lambda I - \mathcal{T}_B^{\delta k}\right)\fbf = \gbf, 
\end{align*}
then, for $\lambda \sim \lambda_{n_0}$,  the following holds:
\begin{align*}
\langle \fbf, \phibf_{n_0}\rangle_{L^2(B,\R^3)} \sim \frac{\langle \gbf, \phibf_{n_0}\rangle_{L^2(B,\R^3)}}{ \lambda - \lambda_{n_0} + \left\langle\left(\mathcal{T}^0 -\mathcal{T}^{\delta k}\right)[\phibf_{n_0}], \phibf_{n_0}\right\rangle_{L^2(B,\R^3)}}.
\end{align*}
\end{proposition}
\proof The result follows directly from Lemma \ref{lem:decomposition} and  identity (\ref{eq:decompf}) with $\gbf$ in the right-hand side.\cqfd
\section{Dipolar resonance of a finite sized particle}\label{sec:finitesize}

\subsection{Computation of the perturbation via change of variables}


By the change of variables: $y=z_0+\delta \tilde{y}$, $\tilde{\Ebf}(\tilde{y})= \Ebf(y)$, $\tilde{\Ebf}^i(\tilde{y}) = \Ebf^i(y)$, and (\ref{eq:fieldinside}) becomes:
\begin{align*}
\left(\frac{\varepsilon_c}{\varepsilon_m - \varepsilon_c} I -  \mathcal{T}^{\delta k}_B\right)  \tilde{\Ebf} = \frac{\varepsilon_c}{\varepsilon_m - \varepsilon_c}\Ebf^i.
\end{align*}

We are now exactly in the right frame to apply the results of Section \ref{sec:perturbative}. 
We know that there is a neighborhood $V_i\subset \mathbb{C}$  of $\lambda_{i}^{(0)}$, $i\in \{1,2,3\}$ such that $\mathcal{T}_B^{\delta k_m}$ has exactly one eigenvalue in $V_i$ (Lemma \ref{lem:oneeigenvalue}) and that the perturbed eigenvalue is given by
\begin{theorem}\label{theo:spectrumperturbation} We have \begin{align}\label{eq:perturbatedev}
\tilde{\lambda}\sim \lambda_{i}^{(0)} - \left\langle\left( \mathcal{T}_B^0 - \mathcal{T}_B^{\delta k_m}\right) \phibf_{i}, \phibf_{i}\right\rangle_{L^2(B,\R^3)}, 
\end{align}
where $\phibf_i$ is a unitary eigenvector of $\mathcal{T}_B^0$ associated with $\lambda_i^{(0)}$.
\end{theorem}

\subsection{The case of an ellipse}
\subsubsection{The perturbative matrix}
In the case where $B$ is an ellipse, since the eigenmodes associated with a dipole resonance are constant $\phibf_i \equiv \Ebf_i \in \R^3$ (see Section \ref{sec:staticellipse}) and therefore formula \eqref{eq:perturbatedev} simplifies to: 
\begin{proposition} We have
\begin{align*}
\left\langle\left(\mathcal{T}_B^0 -\mathcal{T}_B^{\delta k_m}\right)[\phibf_i], \phibf_i\right\rangle =  \Ebf_i\cdot \mathbf{M}^{\delta k_m}_B \Ebf_i
\end{align*}
with
\begin{align*}
\mathbf{M}^{\delta k_m}_B :=\iint_{\partial B \times \partial B} \left( G^0(\tilde{x},\tilde{z}) -G^{\delta k_m} (\tilde{x},\tilde{z})\right) \nubf(\tilde{x})\nubf(\tilde{z})^\top \dd \sigma(\tilde{x}) \dd \sigma(\tilde{z}). 
\end{align*}
\end{proposition}
\proof From
\begin{align*}
\left\langle\left(\mathcal{T}_B^0 -\mathcal{T}_B^{\delta k_m}\right)[\phibf_i], \phibf_i\right\rangle = \Ebf_i\cdot \int_B \nabla\int_B \nabla \left[ G^0(x,y) - G^{\delta k_m}(x,y)\right] \dd y \dd x \Ebf_i,
\end{align*}
an integration by parts yields the result. \cqfd

\subsubsection{The algorithmic procedure}

We now give a practical way to compute this perturbation in the case of an elliptical particle:

\begin{enumerate}
\item[(i)] Compute the resonant value associated with the static problem:
\begin{itemize}
\item Compute the matrix $\mathbf{L}_B\in M_3(\R)$;
\item Compute its spectrum $\lambda_1$, $\lambda_2$, $\lambda_3$ and corresponding unitary eigenvectors $\Ebf_1$, $\Ebf_2$ and $\Ebf_3$;
\end{itemize}
\item[(ii)] Compute the perturbative matrix $\mathbf{M}_B^{\delta k_m}$ and the perturbed eigenvalues $$\tilde{\lambda}_i = \lambda_i - \Ebf_i\cdot \mathbf{M}_B^{\delta k_m}\Ebf_i.$$
\end{enumerate}

\section{Computation of observables for an elliptical nanoparticle}\label{sec:observableelliptic}

\subsection{Dipole moment beyond the quasi-static approximation}
Denote by $\lambda_j$, $ j=1, 2, 3,$ the three eigenvalues of $\mathbf{L}_B$. Denote by $\Ebf_j$ the three eigenvectors ($\in \R^3$) associated with $\lambda_j$ such that $(\Ebf_1,\Ebf_2, \Ebf_3)$ forms an orthonormal basis of $\R^3$.
Denote by $\mathbf{Q}=\left(\Ebf_1, \Ebf_2,\Ebf^3\right)\in O(3)$ the matrix associated with this basis.

Since $\Ebf_j$ are eigenmodes for $\mathcal{T}^0$, we can use Lemma \ref{prop:approxresolv} to find that

\begin{align*}
\langle \Ebf, \Ebf_j \rangle \sim \frac{\langle \Ebf^i, \Ebf_j\rangle}{ \lambda - \lambda_{j} + \left\langle\left(\mathcal{T}^0 -\mathcal{T}^{\delta k}\right)[\Ebf_{j}], \Ebf_{i}\right\rangle}.
\end{align*}
We can then write:

\begin{equation} \label{eq:dipolemoment}\resizebox{.9\hsize}{!}{$
\ds \mathbf{P}\sim \delta^3 \varepsilon_0(\varepsilon -1) \mathbf{Q}\begin{pmatrix}
\ds \frac{1}{ \lambda - \lambda_{1} + \left\langle\left(\mathcal{T}^0 -\mathcal{T}^{\delta k}\right)[\Ebf_{1}], \Ebf_{1}\right\rangle}  &  0 & 0 \\
\nm
0 & \ds \frac{1}{ \lambda - \lambda_{2} + \left\langle\left(\mathcal{T}^0 -\mathcal{T}^{\delta k}\right)[\Ebf_{2}], \Ebf_{2}\right\rangle} & 0 \\
\nm
0 & 0 & \ds \frac{1}{ \lambda - \lambda_{3} + \left\langle\left(\mathcal{T}^0 -\mathcal{T}^{\delta k}\right)[\Ebf_{3}], \Ebf_{3}\right\rangle} & 0
\end{pmatrix} \ds \mathbf{Q}^t \left( \frac{1}{\vert D\vert}\int_D \Ebf^i\right).$}
\end{equation}

\begin{remark}
This expression is valid near the resonant frequencies when the corresponding mode is excited, i.e., when $\lambda \sim \lambda_i $ and $\nabla u^i(z_0) \cdot \Ebf_i \sim \vert \nabla u^i(z_0)\vert $.
\end{remark}
\begin{remark} \label{rem:mdyn}
The expression  $$\resizebox{.9\hsize}{!}{$ \mathcal{M}_{\text{dyn}}:=\delta^3 \varepsilon_0(\varepsilon -1) \mathbf{Q}\begin{pmatrix}
\frac{1}{ \lambda - \lambda_{1} + \left\langle\left(\mathcal{T}^0 -\mathcal{T}^{\delta k}\right)[\Ebf_{1}], \Ebf_{1}\right\rangle}  &  0 & 0 \\
0 & \frac{1}{ \lambda - \lambda_{2} + \left\langle\left(\mathcal{T}^0 -\mathcal{T}^{\delta k}\right)[\Ebf_{2}], \Ebf_{2}\right\rangle} & 0 \\
0 & 0 & \frac{1}{ \lambda - \lambda_{3} + \left\langle\left(\mathcal{T}^0 -\mathcal{T}^{\delta k}\right)[\Ebf_{3}], \Ebf_{3}\right\rangle} & 0
\end{pmatrix} \mathbf{Q}^t$} $$ is a dynamic version of the usual quasi-static polarization tensor. 
\end{remark}

\subsection{Far-field expansion}
Assume that the incident fields are plane waves given by
\begin{align*}
\Ebf^i(x) = \Ebf_0^i e^{i k_m \mathbf{d}\cdot x}, \qquad \Hbf^i(x) =  \mathbf{d}\times \Ebf_0^i e^{ik_m \mathbf{d}\cdot x},
\end{align*} with $\mathbf{d}\in \mathbb{S}^2$ and $\Ebf_0^i\in \R^3$, such that $\Ebf_0^i\cdot \mathbf{d}= 0$.

Since we have an approximation of the dipole moment of the particle we can find an approximation of the electric field radiated far away from the particle. The far-field expansion written in \cite{ammari2016mathematicalruiz} is still valid (Theorem (4.1) in the aforementioned paper), one just has to replace the dipole moment $M(\lambda,D) \Ebf^i$ where $M$ is the usual polarization tensor defined with the Neumann Poincar\'e operator by the new corrected expression obtained in  (\ref{eq:dipolemoment}).

The scattered far field has the form
\begin{align}
\Ebf(x)-\Ebf^i(x) \sim \frac{k_m^2}{4\pi} \frac{e^{ik_m \vert x\vert}}{\vert x \vert} \mathcal{M}_{dyn}  \frac{1}{\vert D \vert}\int_D \Ebf^i(y) \dd y \qquad (\vert x \vert \rightarrow \infty),
\end{align}
and the scattering amplitude is given by
\begin{align*}
\frac{k_m^2}{4\pi}\mathcal{M}_{dyn}  \frac{1}{\vert D \vert}\int_D \Ebf^i(y) \dd y = \frac{k_m^2}{4\pi} \mathbf{P}.
\end{align*}
%

%
%

\subsection{Scattering and absorption cross sections}
Having an approximation of the dipole moment, we can compute the extinction and scattering cross sections of the particle.

\begin{proposition}(\cite[Chapter 13]{born2013principles})
The power radiated by an oscillating dipole $\mathbf{P}$ can be written
\begin{align*}
P_{r} = \frac{\mu_m \omega^4}{12 \pi c_0} \vert \mathbf{P}\vert^2,
\end{align*}
and the power removed from the incident plane wave (absorption and scattering) can be written
\begin{align*}
P_e=\frac{4 \pi }{k_m }\mathcal{I}\left[ \frac{\Ebf^i_0 \cdot  \frac{k_m^2}{4\pi}\mathbf{P}}{\vert \Ebf^i_0 \vert^2}\right].
\end{align*}
\end{proposition}

In the same spirit as in \cite{ammari2015mathematical,ammari2016mathematicalruiz} we can then give upper bounds for the cross sections as follows. 
\begin{proposition}Near plasmonic resonant frequencies, the leading-order term of the average over the orientation of the extinction (respectively absorption) cross section of a randomly oriented nanoparticle is bounded by
\begin{align*}
Q^{ext}_m\sim& k_m \mathcal{I} \left[ \mathrm{Tr}\mathcal{M}_{\mathrm{dyn}}\right], \\
Q^{a}_m\sim& \frac{k_m^4}{6\pi}   \left\vert \mathrm{Tr} \mathcal{M}_{dyn} \right\vert^2, 
\end{align*}
where $\mathrm{Tr}$ denotes the trace. 
\end{proposition}
\proof  We start from equation \eqref{eq:dipolemoment} and get that 
\begin{align*}
\mathbf{P}= M_{dyn} \Ebf^i_0 \left[1+ f(D,k_m,\mathbf{d})\right]
\end{align*}
with $$f(D,k_m,d) = \frac{1}{\vert D \vert}\int_D\left(1-e^{i k_m \dbf\cdot x} \right) \dd x .$$

Here, $f$ represents the correction of the average illuminating field over the particle due to the finite ratio between the size of the particle and the wavelength. Its magnitude is of the order of $\delta k$.
If we take the average over all directions for $\Ebf^i_0$ and $\mathbf{d}$, then we obtain that
\begin{align*}
Q^{ext}_m= &\frac{1}{(4\pi)^2 }\iint_{\mathbb{S}^2}  \frac{4 \pi}{k_m}\frac{k_m^2}{4\pi} \mathcal{I} \left[\ebf_0\cdot \mathcal{M}_{dyn}\ebf_0 (1+f(D,k_m,\mathbf{d}) \right]\dd \sigma(\ebf_0) \dd \sigma(\mathbf{d}).
\end{align*}
The term $f(D,k_m,\dbf)$ is a small correction of the order of $k_m\delta$ that is due to the fact that what determines the dipole moment is not the incident field at the center of the particle, but the average of the field over the particle. Therefore, it reduces the dipole response of the particle. Ignoring it and considering only the leading order term gives:
\begin{align*}
Q^{ext}_m\sim& \frac{k_m}{4\pi} \int_{\mathbb{S}^1}  \mathcal{I} \left[\ebf_0\cdot M_{dyn}\ebf_0\right]  \dd \sigma(\ebf_0)  \\ 
\sim & k_m \mathcal{I} \left[ \mathrm{Tr}\mathcal{M}_{\mathrm{dyn}}\right].
\end{align*}
A similar computation gives the leading term of the absorption cross section.

%
%



\appendix
\section{Justification of the asymptotic regime}
\label{sec:justif}
To quickly justify the model and the regime we are working in, we give some values for the physical parameters used in the model corresponding to practical situations.

In practice: $\omega \in [2, 5]\cdot 10^{15} Hz$;
 $\delta \in [5,100] \cdot 10^{-9}m$;
 $\varepsilon_m\sim 1.8 \varepsilon_0\sim 1.5 \cdot 10^{-11} F\cdot m^{-1}$ for water;
 $\mu_0 \sim 12 \cdot 10^{-7} H\cdot m^{-1}$;
 $c_0\sim 3\cdot 10^8 m\cdot s^{-1}$;
 $k_m \sim 10^7 m^{-1}$.

Therefore, one can see that we have  $\delta k \leq 10^{-2}$ for very small particles, and $\delta k \sim 1$ for bigger  particles (of size $100nm$).
 
For the permittivity of the metal, one can use a Lorentz-Drude type model:
\begin{align*}
\varepsilon_c(\omega)=\varepsilon_0\left(1-\frac{\omega_p^2}{\omega(\omega+i\tau^{-1})}\right) 
\end{align*} with
\begin{itemize}
\item $\tau = 10^{-14}\, s$; 
\item $\omega_p = 2 \cdot 10^{15} s^{-1}$.
\end{itemize}
This model is enough to understand the behavior of $\varepsilon$ but for numerical computations, it is better to use the tabulated parameters that can be found in \cite{rakic1998optical}.
We plot $f(\omega)$ on Figure $2$ and $\frac{\dd f}{\dd \omega}$ on Figure $3$. One can see that $\frac{\dd f}{\dd \omega}$ is of order $10^{-15}$ while $\frac{\delta}{c}\sim 10^{-16}$  for size particle under $100 nm$. So the procedure described in Section \ref{sec:maincontrib} is justified.


%
%
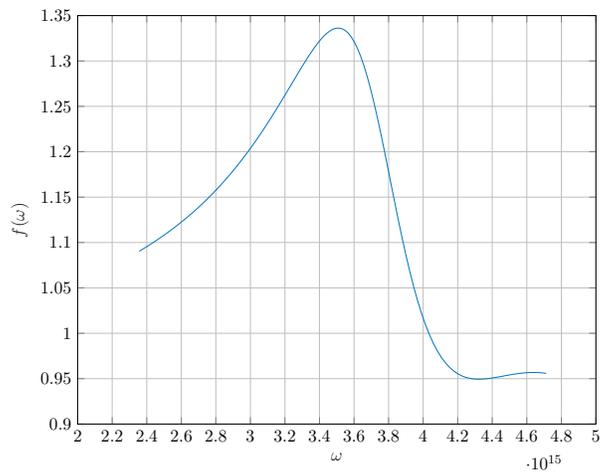
\begin{figure}\label{fig:lambda}

\begin{center}
\input{reallambda.tex}
\caption{Numerical values of the real part of $f(\omega)=\lambda(\omega)= \frac{\varepsilon_c(\omega)}{\varepsilon_m-\varepsilon_c(\omega)}$ for a gold nanoparticle in water.}
\end{center}
\end{figure}

\begin{figure}\label{fig:dlambda}

\begin{center}
\input{realdlambda.tex}
\caption{Numerical values of the real part of $ \frac{\dd f}{\dd \omega}(\omega)$ for a gold nanoparticle in water.}
\end{center}
\end{figure}
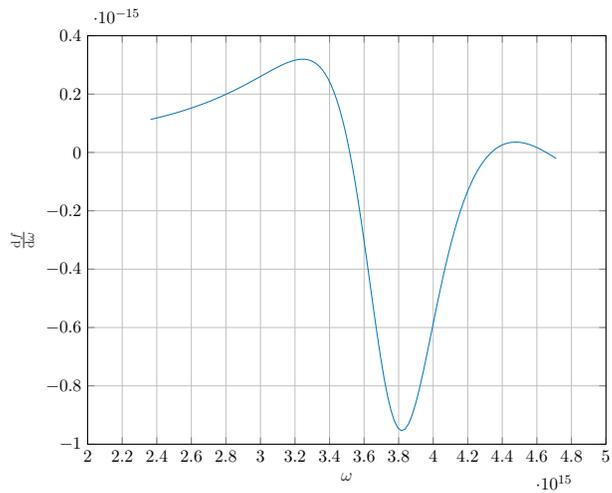

%

\section{Singular integrals, Calder\'on Zygmund type operators}\label{appendix:singular}
There is an abundant literature on singular integral operators, yet these types of \emph{principal value} integrals are misunderstood and misused in some of the physics literature. We include here some properties that are well known for people who are familiar with these types of operators, but seem to be often misunderstood.

There have been numerous contributions in the twentieth century. Some notable ones are: Tricomi ($1928$) \cite{tricomi1928equazioni}; Kellogg ($1929$) \cite{kellogg2012foundations};
Calder\'on-Zygmund ($1952$) \cite{calderon1952existence};
 Seeley ($1959$) \cite{seeley1959singular};
Gel'fand-Shilov ($1964$) \cite{gelfandshilov1964}, and 
Mikhlin ($1965$) \cite{mikhlin2014multidimensional}.

In the following, we do not state the results in their most general settings and assumptions. We use some notations and hypotheses that are adapted to our problem (Green's function method).
\subsection{Principal value integral}
Let $D \subset \R^d$ be a bounded domain. We are concerned with the existence and the  manipulation of integrals of the type

\begin{align}\label{eq:principalvalue}
\int_D f\left(\frac{x-y}{\vert x-y \vert}\right) \frac{1}{\vert x - y \vert^d } u(y) \dd y,\qquad x\in D,
\end{align}
where $u$ is a function defined on $D$ and $f$ a function defined on $\mathbb{S}^{d-1}$. We denote by $B(x,\varepsilon)$ the ball centered at $x$ of radius $\varepsilon$.

\begin{definition}
The principal value of the integral (\ref{eq:principalvalue}) is defined by
\begin{align*}
\lim_{\varepsilon \rightarrow 0} \int_{D\setminus B(x,\varepsilon)} f\left(\frac{x-y}{\vert x-y \vert}\right) \frac{1}{\vert x - y \vert^d } u(y) \dd y.
\end{align*}
\end{definition}

We now give sufficient conditions for the existence of the principal value.
\begin{theorem} \label{theo:pv}
If $u\in \mathcal{C}^{0,\alpha}(D)$, $\alpha>0$ and  $\int_{\mathbb{S}^{d-1}} f(\theta) \dd \theta = 0$,  then
the principal value of (\ref{eq:principalvalue}) does exist. 
\end{theorem}
\begin{remark}These conditions are not necessary, and singular integrals can be defined for a much larger class of functions. $f$ can be replaced by $f(x,\theta)$ and $u$ does not need to be chosen as H\"older continuous. One can choose $u$ in some Lebesgue space $u\in L^p(D)$.
\end{remark}

\begin{example}\label{example:dijG}Consider the Green function for the free space Laplace equation in two and three dimensions: \begin{align*}
G(x,y) = \left\{\begin{aligned} 
&\frac{1}{2\pi}\log \vert x-y\vert \quad \text{if}\quad & d=2 ,\\
&\frac{1}{4\pi} \frac{1}{\vert x-y\vert} \quad  \text{if}\quad & d=3, 
\end{aligned}\right.
\end{align*}
then $\partial_{x_i,x_j} G(x,y) = f\left(\frac{x-y}{\vert x-y \vert}\right) \frac{1}{\vert x - y \vert^d }$ with
\begin{align*}
f\left(\frac{x-y}{\vert x-y \vert}\right) = \left\{\begin{aligned} 
& \frac{1}{2\pi}\left( \delta_{ij} -2 \frac{(x_i-y_i)(x_j-y_j)}{\vert x-y\vert^2}\right)\quad \text{if}\quad & d=2 ,\\
&-\frac{1}{4\pi}\left( \delta_{ij}-3 \frac{(x_i-y_i)(x_j-y_j)}{\vert x-y\vert^2}\right)  \quad  \text{if}\quad & d=3 .
\end{aligned}\right.
\end{align*}
One can check that $\int_{\mathbb{S}^{d-1}} f(\theta) \dd \theta = 0$. Therefore, for $u\in C^{0,\alpha}(D)$  one can write:
\begin{align*}
\int_D \partial_{x_i,x_j} G(x,y) u(y) \dd y = \lim_{\varepsilon \rightarrow 0} \int_{D\setminus B(x,\varepsilon)} \partial_{x_i,x_j} G(x,y) u(y) \dd y.
\end{align*}
\end{example}

\subsection{Non spherical volume of exclusion}
One common misconception found in the physics literature is that the limit of the integral over the domain minus a small volume around the singularity does not depend on the shape of the volume when the maximum cord of the volume of exclusion goes to zero. The limit does depend on the shape of the volume. This issue has been dealt with by Mikhlin \cite[p. 40]{mikhlin2014multidimensional}.
We include here the formula for the limit, using the notations used in physics literature.
Assume that $V(x,\varepsilon)\subset D$ is a small volume of exclusion such that its boundary is given, in polar coordinates by:
\begin{align*}
\partial V(x,\varepsilon) = \left\{ y \in D, \vert x-y\vert = \varepsilon \beta\left( \frac{x-y}{\vert x-y \vert}\right)\right\}.
\end{align*}

\begin{theorem}\label{theo:nonspherical}Under the assumptions of Theorem \ref{theo:pv},
\begin{multline*}
\lim_{\varepsilon \rightarrow 0} \int_{D \setminus V(x,\varepsilon)} f\left(\frac{x-y}{\vert x-y \vert}\right) \frac{1}{\vert x - y \vert^d } u(y) \dd y = \int_D f\left(\frac{x-y}{\vert x-y \vert}\right) \frac{1}{\vert x - y \vert^d } u(y) \dd y  \\ 
- u(x) \int_{\mathbb{S}^{d-1}} f(\theta) \log \beta(\theta) \dd \theta .
\end{multline*}
\end{theorem}
\begin{example} Let $d=2$ and let \begin{align*}
f\left(\frac{x-y}{\vert x-y \vert}\right) = \frac{1}{2\pi} \left(1 -2 \frac{(x_1 - y_1)^2}{\vert x-y\vert^2}\right)
\end{align*} be corresponding to the angular term of $\partial_{1,1}G(x,y)$. If $V(x,\varepsilon)$ is an ellipse of semi-axis $\varepsilon$ and eccentricity $e$ where $x$ is at one of the focal point 
\begin{align*}
\partial V(x,\varepsilon) = \left\{y \in D, \vert y-x\vert = \varepsilon  \frac{(1-e^2)}{1 - e \frac{x_1-y_1}{\vert x-y\vert}} \right\},
\end{align*}
then the correction term is 
\begin{align*}
\frac{u(x)}{2\pi} \int_{\theta \in \mathbb{S}^1} \left(1-2 \theta_1^2\right) \log\left( \frac{1-e^2}{1-e \theta_1} \right)\dd \theta = \frac{u(x)}{2\pi} \int_{0}^{2\pi} (1-2\cos^2(t)) \log\left(\frac{1-e^2}{1-e\cos(t)}\right) \dd t.
\end{align*}
\end{example}
\begin{remark}Note that the correction term does not only depend on the shape of the volume of exclusion, but also on the position of $x$ inside it.
In the previous example, if $x$ is at the center of the ellipse instead of being one of the focal point, the polar equation, hence the correction term, is modified.
\end{remark}

\subsection{Change of variables}
This issue has also been dealt with by Seeley \cite{seeley1959singular} and Mikhlin \cite[p. 41]{mikhlin2014multidimensional}. The classical formula for a change of variables in an integral cannot be applied in a straightforward way, and some precautions have to be taken into account.
Consider a region $\widetilde{D}$ and an homeomorphism $\psibf : D \longrightarrow \widetilde{D}$.
Consider $\tilde{f}= f \circ \psibf^{-1}$, $\tilde{u}= u \circ \psibf^{-1}$, and denote by $\mathbf{J}(\tilde{x})$ the non vanishing Jacobian of $\psibf^{-1}$. The corrective term to the usual change of variables formula is given by the reciprocal image of the unit sphere by $\psibf$. One can establish formulae of the form:
\begin{align*}
\vert x - y \vert^2 = \vert \psibf(x) - \psibf(y) \vert^2 F\left(\psibf(x), \frac{ \psibf(x) - \psibf(y) }{\vert \psibf(x) - \psibf(y) \vert} \right) +O\left(  \vert \psibf(x) - \psibf(y) \vert^3\right),
\end{align*}  
and then the change of variables can be written as follows:
\begin{theorem}\label{theo:changeofvariable} We have

\begin{multline*}
\int_D  f\left(\frac{x-y}{\vert x-y \vert}\right) \frac{1}{\vert x - y \vert^d } u(y) \dd y = \int_{\widetilde{D}}\tilde{f}\left(\frac{\tilde{x}-\tilde{y}}{\vert \tilde{x} - \tilde{y} \vert }\right) \frac{1}{\vert \tilde{x} - \tilde{y} \vert^d } \tilde{u}(\tilde{y}) \mathbf{J}(\tilde{y}) \dd \tilde{y}  \\ + \tilde{u}(\tilde{x}) \mathbf{J}(\tilde{x}) \int_{\mathbf{S}^{d-1}} \tilde{f}(\tilde{\theta}) \log F ( \tilde{x}, \tilde{\theta}) \dd \tilde{\theta}.
\end{multline*}

\end{theorem}
\begin{remark} For a dilation : $\tilde{x}= \frac{x-z_0}{\delta}$, the image of the unit sphere is still a sphere and therefore, $F=1$ and the usual change of variables formula is valid.
\end{remark}

\subsection{Differentiation of weakly singular integrals, integration by parts}
We want to differentiate integrals of the type
\begin{align*}
\int_D g\left( \frac{x-y}{\vert x-y\vert }\right) \frac{1}{\vert x-y\vert^{d-1}} u(y) \dd y,  \qquad x\in D.
\end{align*}
The following results can be found in \cite{seeley1959singular,mikhlin2014multidimensional}:
\begin{theorem}\label{theo:diff} If $u$ is H\"older continuous and if $g$ and its first derivative are bounded,  then: \begin{enumerate}
\item[(i)] Differentiation formula under the integral sign: \begin{multline*}
\frac{\partial}{\partial {x_i}} \int_D g\left( \frac{x-y}{\vert x-y\vert }\right) \frac{1}{\vert x-y\vert^{d-1}} u(y) \dd y = \int_D \frac{\partial}{\partial {x_i}} \left[ g\left( \frac{x-y}{\vert x-y\vert }\right) \frac{1}{\vert x-y\vert^{d-1}} \right] u(y) \dd y \\ + u(x) \int_{\mathbb{S}^{d-1}} g(\theta)  \theta_i \dd \theta;
\end{multline*}
\item[(ii)] Integration by parts formula:\begin{multline*}
\int_D g\left( \frac{x-y}{\vert x-y\vert }\right) \frac{1}{\vert x-y\vert^{d-1}}  \frac{\partial}{\partial {x_i}} \left[f(y)\right] \dd y =  - \int_D  \frac{\partial}{\partial {x_i}} \left[g\left( \frac{x-y}{\vert x-y\vert }\right) \frac{1}{\vert x-y\vert^{d-1}}\right]  f(y) \dd y  \\ + \int_{\partial D} g\left( \frac{x-y}{\vert x-y\vert }\right) \frac{f(y)}{\vert x-y\vert^{d-1}} \nubf(y) \cdot \ebf_i \dd \sigma(y)+ f(x) \int_{\mathbb{S}^{d-1}} g(\theta)  \theta_i \dd \theta.
\end{multline*}
\end{enumerate}
\end{theorem}

\begin{remark} Once again we only give sufficient conditions for the validity of theses formulas, corresponding to our framework. These formulas are valid for $u\in L^p$ and for more general kernels. 
\end{remark}


\begin{example}
Let $d=3$ and consider the second derivative of a Newtonian potential:
\begin{align*}
\partial_{x_i,x_j} \int_D \frac{1}{4\pi} \vert x-y\vert^{-1} u(y) \dd y= -\partial_{x_i} \int_D\frac{1}{4\pi} \frac{x_j-y_j}{\vert x-y\vert^3}  u(y) \dd y.
\end{align*}
We can apply Theorem \ref{theo:diff} with $g(\theta) = \theta_j$ and get:
\begin{align*}
\partial_{x_i,x_j} \int_D\frac{1}{4\pi}  \vert x-y\vert^{-1} u(y) \dd y= \int_D \frac{1}{4\pi}\partial_{x_i,x_j} \left[ \vert x-y\vert^{-1}\right]  u(y)\dd y + u(x) \frac{\delta_{ij}}{3}.
\end{align*}
\end{example}

\subsection{The $\mathbf{L}$ dyadic}
\begin{lemma}\label{lem:L_v} Let $x\in D$.
Denote by $\mathbf{L}_D(x)$ the matrix
\begin{align*}
\int_{\partial D}\nabla G(x,y) \nubf^\top(y)\dd \sigma (y).
\end{align*} Assume that $D$ can be written in polar coordinates as
\begin{align*}
D=\left\{y\in \R^d, \vert x-y\vert \leq \rho\left(\frac{x-y}{\vert x-y\vert}\right) \right\}.
\end{align*}
Then, 
\begin{align*}
\left(\mathbf{L}_D(x)\right)_{i,j}= \int_{\partial D} \partial_{x_j} G(x,y) \nubf(y) \cdot \ebf_i \dd \sigma(y) = -\frac{1}{d} + \int_{\theta\in \mathbb{S}^{d-1}} f_{i,j}(\theta) \log \rho(\theta) \dd \theta 
\end{align*}
with $f_{i,j}$ being defined in Example \ref{example:dijG}.
\end{lemma}

\proof
We start by using the integration by part formula from Theorem \ref{theo:diff}, with $g(\theta) = \frac{1}{2\pi } \theta_j$ if $d=2$ and  $g(\theta)=\frac{1}{4\pi} \theta_j$ if $d=3$, and $f=1$. We obtain
\begin{align*}
\int_D \partial_{x_i,x_j}G(x,y) \dd y= \int_{\partial D} \partial_{x_j}G(x,y) \nubf(y)\cdot\ebf_i \dd \sigma(y)  + \frac{1}{d}.
\end{align*}

In order to compute $\int_D\partial_{x_i,x_j}G(x,y)\dd y$, we use the change of variables \cite{seeley1959singular,mikhlin2014multidimensional} : $y=x+t \theta$, $\theta\in \mathbb{S}^{d-1}$ and $t\in [0,\rho(\theta)]$ to arrive at
\begin{align*}
\int_{D} \partial_{x_i,x_j}G(x,y) \dd y =&  \int_{\theta\in \mathbb{S}^{d-1}} \int_{t\in [0,\rho(\theta)]} f_{i,j}(\theta) t^{-d} t^{d-1} \dd t \dd \theta \\=&\int_{\theta\in \mathbb{S}^{d-1}} f_{i,j}(\theta) \log \rho(\theta) \dd \theta.
\end{align*}
\cqfd

\subsection{Second derivative of a Newtonian potential}
In this section we give a correct simple derivation of the formula found in \cite[p. 73]{van1991singular}  and \cite{yaghjian1980electric,lee1980singularity,asvestas1983comments}.
\begin{proposition}
Let $V^*\subset \R^d$ be such that \begin{enumerate}
\item[(i)] $0\in V^*$;
\item[(ii)] $\partial V^*$ is a piecewise smooth;
\item[(iii)] $V^*$ is radially convex with respect to the origin.
\end{enumerate} Let $V(x,\varepsilon)= x+ \varepsilon V^*$. Then,  
\begin{align*}
\partial_{x_i,x_j}\int_D G(x,y) u(y) \dd y = \lim_{\varepsilon \rightarrow 0} \int_{D\setminus V(x,\varepsilon)} \partial_{x_i,x_j} G(x,y) u(y)\dd y  - \left(\mathbf{L}_{V^*}\right)_{ij} u(x).
\end{align*}
\end{proposition}

\proof
Let $x\in D$ and let $V(x,\varepsilon)\subset D$. 
Assume that $V(x,\varepsilon)$ can be described by some polar equation:
\begin{align}\label{eq:eqVeps}
V(x,\varepsilon)= \left\{y\in D, \vert x - y \vert \leq \varepsilon \rho\left(\frac{x-y}{\vert x-y\vert}\right) \right\}.
\end{align}
Before the computation we also recall that $\partial_{x_i,x_j} G(x,y)$ can be written as \begin{align*}
\partial_{x_i,x_j} G(x,y) = f_{ij}\left(\frac{x-y}{\vert x-y\vert} \right)\vert x-y\vert^{-d},
\end{align*} as it was seen in Example \ref{example:dijG}. Then we have
\begin{align}\label{eq:split1}
\partial_{x_i,x_j}\int_D G(x,y) u(y) \dd y =& \partial_{x_i} \int_D \partial_{x_j} G(x,y) u(y)\dd y 
\\ =& \frac{1}{d} u(x) + \int_D\partial_{x_i,x_j}G(x,y) u(y) \dd y .
\end{align}
Using Theorem \ref{theo:nonspherical} we obtain:
\begin{align*}
\int_D\partial_{x_i,x_j}G(x,y) u(y) \dd y  = \lim_{\varepsilon \rightarrow 0 } \int_{D\setminus V(x,\varepsilon)} \partial_{x_i,x_j}G(x,y) u(y) \dd y  - u(x) \int_{\mathbb{S}^{d-1}} f_{ij}(\theta) \log \rho(\theta) \dd \theta, 
\end{align*}
and therefore,
\begin{align*}
\partial_{x_i} \int_D \partial_{x_j} G(x,y) u(y)\dd y = \lim_{\varepsilon \rightarrow 0 }\int_{D\setminus V(x,\varepsilon)} \partial_{x_i,x_j} G(x,y) u(y)\dd y    - u(x) \left(- \frac{1}{d} + \int_{\theta\in \mathbb{S}^{d-1}} f_{i,j}(\theta) \log \rho(\theta) \dd \theta\right).
\end{align*}
Finally, using Lemma \ref{lem:L_v} we arrive at
\begin{align*}
\partial_{x_i} \int_D \partial_{x_j} G(x,y) u(y)\dd y = \lim_{\varepsilon \rightarrow 0 }\int_{D\setminus V(x,\varepsilon)} \partial_{x_i,x_j} G(x,y) u(y)\dd y    - \left(\mathbf{L}_{V^*}\right)_{ij} u(x).
\end{align*} \cqfd
\begin{remark} There are several issues and misconceptions in the literature with this formula:
\begin{enumerate}
\item[(i)] The shape $V^*$ cannot be completely arbitrary as often mentioned. It has to satisfy some regularity condition, since the construction of $\mathbf{L}_{V^*}$ uses some integration on the boundary of $V^*$ involving the normal vector. 
\item[(ii)] The exclusion volume $V(x,\varepsilon)$ needs to be taken small in the numerical evaluation of the integral. Only if the test function $u$ is constant then $\varepsilon$ does not need to be small.
\item[(iii)] The derivation of this formula often contains mistakes. One common derivation of this formula is through a splitting of the integral of the form
\begin{multline*}
\partial_{x_i} \int_D \partial_{x_j} G(x,y) u(y)\dd y = \int_{D\setminus V(x,\varepsilon)} \partial_{x_i,x_j} G(x,y) u(y)\dd y   \\ +\int_{V(x,\varepsilon)} \partial_{x_i,x_j} G(x,y) \left[u(y)-u(x)\right]\dd y  +   u(x)\partial_{x_i} \left(\int_{V(x,\varepsilon)} \partial_{x_j} G(x,y) \dd y\right),
\end{multline*} which is a wrong application of the differentiation under the $\int$ sign theorem.
The reason why it is wrong is that, if the limit when  $\varepsilon \rightarrow 0$ is to be taken, then one has to take into account the dependency of the volume of integration on $D\setminus V(x,\varepsilon)$ on the variable $x$ and use Reynold's transport theorem to compute the derivative and add some boundary integral terms.
The correct splitting would be:
\begin{multline*}
\partial_{x_i} \int_D \partial_{x_j} G(x,y) u(y)\dd y = \int_{D\setminus V(x,\varepsilon)} \partial_{x_i,x_j} G(x,y) u(y)\dd y  - \int_{\partial V(x,\varepsilon)} \partial_{x_j} G(x,y) u(y) \nubf(y)\cdot \ebf_i \dd \sigma(y) \\ +\int_{V(x,\varepsilon)} \partial_{x_i,x_j} G(x,y) \left[u(y)-u(x)\right]\dd y  + \int_{\partial V(x,\varepsilon)} \partial_{x_j}G(x,y) \left[u(y)-u(x)\right] \nubf(y) \cdot \ebf_i \dd \sigma(y) \\
+\partial_{x_i} u(x) \int_{V(x,\varepsilon)} \partial_{x_j}G(x,y) \dd y  + u(x) \left(\int_{\partial V(x,\varepsilon)} \partial_{x_j}G(x,y) \nubf(y)\cdot\ebf_i\dd \sigma(y) + \frac{1}{d} + \int_{V(x,\varepsilon)} \partial_{x_i,x_j} G(x,y) \dd y\right).
\end{multline*}
In the limit $\varepsilon \rightarrow 0$, the extra terms compensate each other and the first (wrong) splitting gives the same (correct) result as the second one.
\end{enumerate}
\end{remark}

\bibliographystyle{plain}
\bibliography{biblio}

\end{document}

%% file: schema.pdf_tex
\begingroup%
  \makeatletter%
  \providecommand\color[2][]{%
    \errmessage{(Inkscape) Color is used for the text in Inkscape, but the package 'color.sty' is not loaded}%
    \renewcommand\color[2][]{}%
  }%
  \providecommand\transparent[1]{%
    \errmessage{(Inkscape) Transparency is used (non-zero) for the text in Inkscape, but the package 'transparent.sty' is not loaded}%
    \renewcommand\transparent[1]{}%
  }%
  \providecommand\rotatebox[2]{#2}%
  \ifx\svgwidth\undefined%
    \setlength{\unitlength}{841.88976378bp}%
    \ifx\svgscale\undefined%
      \relax%
    \else%
      \setlength{\unitlength}{\unitlength * \real{\svgscale}}%
    \fi%
  \else%
    \setlength{\unitlength}{\svgwidth}%
  \fi%
  \global\let\svgwidth\undefined%
  \global\let\svgscale\undefined%
  \makeatother%
  \begin{picture}(1,0.70707071)%
    \put(0,0){\includegraphics[width=\unitlength,page=1]{schema.pdf}}%
    \put(0.48831284,0.36016689){\color[rgb]{0,0,0}\makebox(0,0)[lb]{\smash{$z_0$}}}%
    \put(0,0){\includegraphics[width=\unitlength,page=2]{schema.pdf}}%
    \put(0.38705614,0.16278863){\color[rgb]{0,0,0}\makebox(0,0)[lb]{\smash{$E^i(\mathbf{x},\omega)$}}}%
    \put(0.64851179,0.41499663){\color[rgb]{0,0,0}\makebox(0,0)[lb]{\smash{$\varepsilon_m$}}}%
    \put(0.53502596,0.41284804){\color[rgb]{0,0,0}\makebox(0,0)[lb]{\smash{$\varepsilon_c(\omega)$}}}%
    \put(0.73305981,0.22570385){\color[rgb]{0,0,0}\makebox(0,0)[lb]{\smash{$D=z_0+\delta B$}}}%
    \put(0,0){\includegraphics[width=\unitlength,page=3]{schema.pdf}}%
    \put(0.30864145,0.60947473){\color[rgb]{0,0,0}\makebox(0,0)[lb]{\smash{$E^s(\mathbf{x},\omega)=\ ?$}}}%
    \put(0,0){\includegraphics[width=\unitlength,page=4]{schema.pdf}}%
    \put(0.44950461,0.41448538){\color[rgb]{0,0,0}\makebox(0,0)[lb]{\smash{}}}%
    \put(0.46113709,0.40503123){\color[rgb]{0,0,0}\makebox(0,0)[lb]{\smash{$\delta$}}}%
    \put(0.73822709,0.19152118){\color[rgb]{0,0,0}\makebox(0,0)[lb]{\smash{$\delta \omega \ll 1$}}}%
  \end{picture}%
\endgroup%

%% file: reallambda.tex
%
%
\definecolor{mycolor1}{rgb}{0.00000,0.44700,0.74100}%
\begin{tikzpicture}[thick, scale=0.6]

\begin{axis}[%
width=4.521in,
height=3.566in,
at={(0.758in,0.481in)},
scale only axis,
xmin=2e+15,
xmax=5e+15,
xlabel style={font=\color{white!15!black}},
xlabel={$\omega$},
ymin=0.9,
ymax=1.35,
ylabel style={font=\color{white!15!black}},
ylabel={$f(\omega)$},
axis background/.style={fill=white},
xmajorgrids,
ymajorgrids
]
\addplot [color=mycolor1, forget plot]
  table[row sep=crcr]{%
4.71238898038469e+15	0.955890908695786\\
4.68097305384879e+15	0.95654853675684\\
4.64997323230012e+15	0.956840141495974\\
4.61938130314026e+15	0.956796734063845\\
4.58918926847921e+15	0.956457985985967\\
4.55938933816441e+15	0.955870868129954\\
4.52997392307948e+15	0.95508839060898\\
4.50093562870076e+15	0.954168516349858\\
4.47226724890012e+15	0.953173290803276\\
4.44396175998303e+15	0.952168203203418\\
4.41601231495169e+15	0.951221772402743\\
4.38841223798324e+15	0.950405332865614\\
4.36115501911378e+15	0.949792983281312\\
4.33423430911925e+15	0.949461650352305\\
4.30764391458478e+15	0.949491212376664\\
4.28137779315438e+15	0.949964620121102\\
4.25543004895345e+15	0.950967945297467\\
4.22979492817662e+15	0.952590279248313\\
4.20446681483424e+15	0.954923396288341\\
4.17944022665071e+15	0.958061088270626\\
4.15470981110839e+15	0.962098070877685\\
4.13027034163129e+15	0.967128360328575\\
4.10611671390245e+15	0.973243025087029\\
4.08224394230999e+15	0.980527235113789\\
4.05864715651629e+15	0.989056566169649\\
4.03532159814551e+15	0.99889257341388\\
4.01226261758468e+15	1.01007773036481\\
3.98946567089386e+15	1.02262993618589\\
3.96692631682101e+15	1.03653692065745\\
3.94464021391752e+15	1.05175100883426\\
3.92260311775038e+15	1.06818482424985\\
3.90081087820733e+15	1.08570858097168\\
3.87925943689126e+15	1.10414960806767\\
3.85794482460065e+15	1.12329463734682\\
3.83686315889245e+15	1.14289515443101\\
3.81601064172456e+15	1.16267577770931\\
3.7953835571747e+15	1.18234523334508\\
3.7749782692329e+15	1.20160910711329\\
3.75479121966481e+15	1.22018325644569\\
3.73481892594318e+15	1.23780662862018\\
3.71505797924507e+15	1.25425229051193\\
3.6955050425122e+15	1.2693357227283\\
3.67615684857235e+15	1.28291981311106\\
3.65701019831937e+15	1.29491642031599\\
3.63806195894984e+15	1.3052847822501\\
3.61930906225422e+15	1.31402734985723\\
3.60074850296061e+15	1.32118379947143\\
3.58237733712918e+15	1.32682401591623\\
3.56419268059553e+15	1.33104076887011\\
3.54619170746121e+15	1.33394266527975\\
3.52837164862974e+15	1.33564779055932\\
3.51072979038659e+15	1.33627828351312\\
3.49326347302149e+15	1.33595594583544\\
3.47597008949168e+15	1.33479887708611\\
3.45884708412472e+15	1.33291905186838\\
3.44189195135941e+15	1.33042071357027\\
3.42510223452351e+15	1.32739944174904\\
3.40847552464718e+15	1.32394175072283\\
3.39200945931072e+15	1.32012508858556\\
3.37570172152557e+15	1.31601812341578\\
3.35955003864746e+15	1.31168122310578\\
3.34355218132057e+15	1.30716705450735\\
3.32770596245175e+15	1.30252124503595\\
3.31200923621377e+15	1.29778306481672\\
3.29645989707661e+15	1.29298609971987\\
3.28105587886598e+15	1.28815889534963\\
3.26579515384799e+15	1.28332555951239\\
3.25067573183944e+15	1.27850631624776\\
3.23569565934248e+15	1.27371800852162\\
3.2208530187033e+15	1.26897454947835\\
3.20614592729369e+15	1.2642873240164\\
3.19157253671509e+15	1.25966554362157\\
3.17713103202407e+15	1.25511655805383\\
3.16281963097891e+15	1.25064612778515\\
3.14863658330636e+15	1.24625866113849\\
3.13458016998803e+15	1.24195741996685\\
3.12064870256586e+15	1.23774469749786\\
3.10684052246601e+15	1.23362197169717\\
3.09315400034061e+15	1.22959003720449\\
3.07958753542684e+15	1.22564911859098\\
3.06613955492279e+15	1.22179896738789\\
3.05280851337965e+15	1.2180389450551\\
3.0395928921096e+15	1.21436809379661\\
3.02649119860913e+15	1.21078519689276\\
3.01350196599708e+15	1.20728883000472\\
3.00062375246717e+15	1.20387740471631\\
2.98785514075455e+15	1.20054920540968\\
2.97519473761576e+15	1.19730242042296\\
2.96264117332202e+15	1.1941351683084\\
2.9501931011652e+15	1.19104551989663\\
2.93784919697623e+15	1.18803151677414\\
2.92560815865549e+15	1.18509118669651\\
2.91346870571502e+15	1.18222255638596\\
2.9014295788319e+15	1.17942366209849\\
2.88948953941283e+15	1.17669255829142\\
2.87764736916934e+15	1.1740273246745\\
2.86590186970334e+15	1.17142607188797\\
2.85425186210292e+15	1.1688869460155\\
2.84269618654785e+15	1.16640813211053\\
2.83123370192467e+15	1.1639878568886\\
2.81986328545108e+15	1.16162439071632\\
2.80858383230928e+15	1.15931604900875\\
2.79739425528812e+15	1.15706119313074\\
2.7862934844338e+15	1.1548582308838\\
2.77528046670877e+15	1.15270561664826\\
2.76435416565874e+15	1.15060185124016\\
2.75351356108752e+15	1.14854548153361\\
2.74275764873953e+15	1.14653509989177\\
2.73208543998957e+15	1.14456934344322\\
2.72149596154e+15	1.14264689323485\\
2.71098825512478e+15	1.14076647328785\\
2.70056137722046e+15	1.13892684957916\\
2.69021439876367e+15	1.13712682896724\\
2.67994640487526e+15	1.13536525807827\\
2.66975649459057e+15	1.13364102216598\\
2.6596437805959e+15	1.13195304395652\\
2.64960738897101e+15	1.13030028248759\\
2.63964645893729e+15	1.12868173194959\\
2.62976014261168e+15	1.12709642053523\\
2.61994760476612e+15	1.12554340930286\\
2.61020802259226e+15	1.12402179105756\\
2.60054058547155e+15	1.12253068925363\\
2.59094449475025e+15	1.121069256921\\
2.58141896351955e+15	1.11963667561763\\
2.57196321640043e+15	1.11823215440952\\
2.56257648933328e+15	1.11685492887937\\
2.55325802937207e+15	1.11550426016474\\
2.54400709448304e+15	1.11417943402596\\
2.53482295334772e+15	1.11287975994429\\
2.52570488517021e+15	1.11160457024995\\
2.5166521794886e+15	1.11035321928006\\
2.50766413599042e+15	1.10912508256613\\
2.4987400643321e+15	1.10791955605048\\
2.48987928396212e+15	1.10673605533122\\
2.48108112394812e+15	1.10557401493502\\
2.47234492280746e+15	1.1044328876171\\
2.46367002834147e+15	1.10331214368762\\
2.45505579747314e+15	1.10221127036375\\
2.44650159608822e+15	1.1011297711466\\
2.43800679887958e+15	1.1000671652223\\
2.42957078919487e+15	1.09902298688633\\
2.42119295888731e+15	1.09799678499036\\
2.41287270816948e+15	1.09698812241076\\
2.40460944547027e+15	1.09599657553804\\
2.3964025872946e+15	1.09502173378643\\
2.38825155808612e+15	1.09406319912283\\
2.38015579009261e+15	1.09312058561434\\
2.37211472323418e+15	1.09219351899379\\
2.36412780497414e+15	1.0912816362424\\
2.35619449019234e+15	1.09038458518895\\
};
\end{axis}
\end{tikzpicture}%

%% file: realdlambda.tex
%
%
\definecolor{mycolor1}{rgb}{0.00000,0.44700,0.74100}%
\begin{tikzpicture}[thick, scale=0.6]

\begin{axis}[%
width=4.521in,
height=3.566in,
at={(0.758in,0.481in)},
scale only axis,
xmin=2e+15,
xmax=5e+15,
xlabel style={font=\color{white!15!black}},
xlabel={$\omega$},
ymin=-1e-15,
ymax=4e-16,
ylabel style={font=\color{white!15!black}},
ylabel={$\frac{\dd f}{\dd \omega}$},
axis background/.style={fill=white},
xmajorgrids,
ymajorgrids
]
\addplot [color=mycolor1, forget plot]
  table[row sep=crcr]{%
4.71238898038469e+15	-2.09329513265282e-17\\
4.68097305384879e+15	-9.40665863757795e-18\\
4.64997323230012e+15	1.41891777737949e-18\\
4.61938130314026e+15	1.12197830216006e-17\\
4.58918926847921e+15	1.9701987548661e-17\\
4.55938933816441e+15	2.66009341943377e-17\\
4.52997392307948e+15	3.16779714099503e-17\\
4.50093562870076e+15	3.47150956385714e-17\\
4.47226724890012e+15	3.55085758384718e-17\\
4.44396175998303e+15	3.38622394689425e-17\\
4.41601231495169e+15	2.95810601565367e-17\\
4.38841223798324e+15	2.24655929585027e-17\\
4.36115501911378e+15	1.23077336769552e-17\\
4.33423430911925e+15	-1.11175576283147e-18\\
4.30764391458478e+15	-1.80235116057062e-17\\
4.28137779315438e+15	-3.86671445731626e-17\\
4.25543004895345e+15	-6.32855981046127e-17\\
4.22979492817662e+15	-9.21157059150161e-17\\
4.20446681483424e+15	-1.253743402526e-16\\
4.17944022665071e+15	-1.63239578411135e-16\\
4.15470981110839e+15	-2.05826458532685e-16\\
4.13027034163129e+15	-2.53157199701007e-16\\
4.10611671390245e+15	-3.05126281569322e-16\\
4.08224394230999e+15	-3.61461562198798e-16\\
4.05864715651629e+15	-4.21683677958618e-16\\
4.03532159814551e+15	-4.85067278729971e-16\\
4.01226261758468e+15	-5.50609078984141e-16\\
3.98946567089386e+15	-6.17009006851166e-16\\
3.96692631682101e+15	-6.82671539420877e-16\\
3.94464021391752e+15	-7.45734160750761e-16\\
3.92260311775038e+15	-8.04128308483951e-16\\
3.90081087820733e+15	-8.55674886219387e-16\\
3.87925943689126e+15	-8.98211471929316e-16\\
3.85794482460065e+15	-9.29742334191696e-16\\
3.83686315889245e+15	-9.48596426946022e-16\\
3.81601064172456e+15	-9.53574199408806e-16\\
3.7953835571747e+15	-9.44062824457869e-16\\
3.7749782692329e+15	-9.20102230380143e-16\\
3.75479121966481e+15	-8.82390997255135e-16\\
3.73481892594318e+15	-8.32230466636344e-16\\
3.71505797924507e+15	-7.71415180360889e-16\\
3.6955050425122e+15	-7.02085705000963e-16\\
3.67615684857235e+15	-6.26564283904546e-16\\
3.65701019831937e+15	-5.47193949364115e-16\\
3.63806195894984e+15	-4.66198249211054e-16\\
3.61930906225422e+15	-3.85572950738857e-16\\
3.60074850296061e+15	-3.0701461717545e-16\\
3.58237733712918e+15	-2.31885213013428e-16\\
3.56419268059553e+15	-1.61207751824141e-16\\
3.54619170746121e+15	-9.5685726724932e-17\\
3.52837164862974e+15	-3.57384661590729e-17\\
3.51072979038659e+15	1.84548162582353e-17\\
3.49326347302149e+15	6.69081760280053e-17\\
3.47597008949168e+15	1.09783602670301e-16\\
3.45884708412472e+15	1.47349969633793e-16\\
3.44189195135941e+15	1.79947753185085e-16\\
3.42510223452351e+15	2.07960026483419e-16\\
3.40847552464718e+15	2.31789565951585e-16\\
3.39200945931072e+15	2.51841501493943e-16\\
3.37570172152557e+15	2.68510739266584e-16\\
3.35955003864746e+15	2.82173325226505e-16\\
3.34355218132057e+15	2.93180948077411e-16\\
3.32770596245175e+15	3.01857861785086e-16\\
3.31200923621377e+15	3.08499612397999e-16\\
3.29645989707661e+15	3.13373063068614e-16\\
3.28105587886598e+15	3.16717313990644e-16\\
3.26579515384799e+15	3.18745204803713e-16\\
3.25067573183944e+15	3.19645163740575e-16\\
3.23569565934248e+15	3.19583230409324e-16\\
3.2208530187033e+15	3.18705128798282e-16\\
3.20614592729369e+15	3.17138305592108e-16\\
3.19157253671509e+15	3.14993878066778e-16\\
3.17713103202407e+15	3.12368457467622e-16\\
3.16281963097891e+15	3.09345829468912e-16\\
3.14863658330636e+15	3.05998484409403e-16\\
3.13458016998803e+15	3.02388997606817e-16\\
3.12064870256586e+15	2.98571265067999e-16\\
3.10684052246601e+15	2.94591603018915e-16\\
3.09315400034061e+15	2.90489721424156e-16\\
3.07958753542684e+15	2.86299582448679e-16\\
3.06613955492279e+15	2.82050154943002e-16\\
3.05280851337965e+15	2.77766075728272e-16\\
3.0395928921096e+15	2.73468227883429e-16\\
3.02649119860913e+15	2.69174245505282e-16\\
3.01350196599708e+15	2.6489895360781e-16\\
3.00062375246717e+15	2.60654751004401e-16\\
2.98785514075455e+15	2.56451943206806e-16\\
2.97519473761576e+15	2.52299031609325e-16\\
2.96264117332202e+15	2.48202964511577e-16\\
2.9501931011652e+15	2.44169354877326e-16\\
2.93784919697623e+15	2.40202669134931e-16\\
2.92560815865549e+15	2.36306390792158e-16\\
2.91346870571502e+15	2.3248316216247e-16\\
2.9014295788319e+15	2.28734907082025e-16\\
2.88948953941283e+15	2.25062937123407e-16\\
2.87764736916934e+15	2.21468043488527e-16\\
2.86590186970334e+15	2.17950576476675e-16\\
2.85425186210292e+15	2.14510514175894e-16\\
2.84269618654785e+15	2.11147521806815e-16\\
2.83123370192467e+15	2.07861002960992e-16\\
2.81986328545108e+15	2.04650143810067e-16\\
2.80858383230928e+15	2.01513951219528e-16\\
2.79739425528812e+15	1.98451285577285e-16\\
2.7862934844338e+15	1.95460889039472e-16\\
2.77528046670877e+15	1.92541409802775e-16\\
2.76435416565874e+15	1.89691422931482e-16\\
2.75351356108752e+15	1.86909448198316e-16\\
2.74275764873953e+15	1.84193965336237e-16\\
2.73208543998957e+15	1.81543427046792e-16\\
2.72149596154e+15	1.78956270064342e-16\\
2.71098825512478e+15	1.76430924536881e-16\\
2.70056137722046e+15	1.73965821948379e-16\\
2.69021439876367e+15	1.71559401779744e-16\\
2.67994640487526e+15	1.69210117078088e-16\\
2.66975649459057e+15	1.66916439083002e-16\\
2.6596437805959e+15	1.64676861037576e-16\\
2.64960738897101e+15	1.62489901296973e-16\\
2.63964645893729e+15	1.60354105830813e-16\\
2.62976014261168e+15	1.58268050204351e-16\\
2.61994760476612e+15	1.56230341111364e-16\\
2.61020802259226e+15	1.54239617523119e-16\\
2.60054058547155e+15	1.52294551507976e-16\\
2.59094449475025e+15	1.5039384877061e-16\\
2.58141896351955e+15	1.48536248951947e-16\\
2.57196321640043e+15	1.46720525726453e-16\\
2.56257648933328e+15	1.44945486728445e-16\\
2.55325802937207e+15	1.43209973333973e-16\\
2.54400709448304e+15	1.41512860323012e-16\\
2.53482295334772e+15	1.39853055441303e-16\\
2.52570488517021e+15	1.38229498880917e-16\\
2.5166521794886e+15	1.36641162693402e-16\\
2.50766413599042e+15	1.35087050149687e-16\\
2.4987400643321e+15	1.33566195058324e-16\\
2.48987928396212e+15	1.32077661051006e-16\\
2.48108112394812e+15	1.30620540844237e-16\\
2.47234492280746e+15	1.29193955484993e-16\\
2.46367002834147e+15	1.27797053585142e-16\\
2.45505579747314e+15	1.26429010551172e-16\\
2.44650159608822e+15	1.2508902781322e-16\\
2.43800679887958e+15	1.23776332056611e-16\\
2.42957078919487e+15	1.22490174460188e-16\\
2.42119295888731e+15	1.21229829942943e-16\\
2.41287270816948e+15	1.19994596421927e-16\\
2.40460944547027e+15	1.1878379408278e-16\\
2.3964025872946e+15	1.17596764664432e-16\\
2.38825155808612e+15	1.16432870759421e-16\\
2.38015579009261e+15	1.15291495130027e-16\\
2.37211472323418e+15	1.14172040041887e-16\\
2.36412780497414e+15	1.13073926614463e-16\\
};
\end{axis}
\end{tikzpicture}%